\documentclass[journal,onecolumn]{IEEEtran}
\usepackage{cite}
\usepackage{url}
\usepackage{amsmath,amssymb,amsfonts}
\usepackage{algorithmic}
\usepackage{graphicx}
\usepackage{subfig}
\usepackage{textcomp}
\usepackage{booktabs}
\usepackage{multirow}
\usepackage{balance}

\newlength{\tempheight}
\newlength{\tempwidth}

\newcommand{\rowname}[1]
{\rotatebox{90}{\makebox[\tempheight][c]{\textbf{#1}}}}

\newcommand{\columnname}[1]
{\makebox[\tempwidth][c]{\textbf{#1}}}

\def\BibTeX{{\rm B\kern-.05em{\sc i\kern-.025em b}\kern-.08em
    T\kern-.1667em\lower.7ex\hbox{E}\kern-.125emX}}

\begin{document}

\title{Artificial Intelligence applied to chest X-Ray images for the automatic detection of COVID-19. A thoughtful evaluation approach}

\author{ Juli\'{a}n D. Arias-Londo\~{n}o\thanks{Juli\'{a}n D. Arias-Londo\~{n}o is with the Department of Systems Engineering, Universidad de Antioquia. Calle 67 No. 53 - 108, 050010, Medell\'{i}n, Colombia. (e-mail: julian.ariasl@udea.edu.co)}, Jorge A. G\'omez-Garc\'ia,  Laureano Moro-Vel\'azquez\thanks{Laureano Moro-Vel\'azquez is with the Department of Electrical and Computer Engineering, Johns Hopkins University, 3400 North Charles St., Baltimore, MD, 21218, USA. (e-mail: laureano@jhu.edu)}, and Juan I. Godino-Llorente\thanks{Jorge A. G\'omez-Garc\'ia and Juan I. Godino-Llorente are with Bioengineering and Optoelectronics lab (ByO). Universidad Polit\'{e}cnica de Madrid. Ctra. Valencia, km. 7, 28031. Madrid, Spain. (e-mails: jorge.gomez.garcia@upm.es, ignacio.godino@upm.es)}
\thanks{This work was supported by the Ministry of Economy and Competitiveness of Spain under grant DPI2017-83405-R1 and Universidad de Antioquia, Medell\'in, Colombia.}}

\maketitle

\begin{abstract}
Current standard protocols used in the clinic for diagnosing COVID-19 include molecular or antigen tests, generally complemented by a plain chest X-Ray. The combined analysis aims to reduce the significant number of false negatives of these tests, but also to provide complementary evidence about the presence and severity of the disease. However, the procedure is not free of errors, and the interpretation of the chest X-Ray is only restricted to radiologists due to its complexity. 
With the long term goal to provide new evidence for the diagnosis, this paper presents an evaluation of different methods based on a deep neural network. 
These are the first steps to develop an automatic COVID-19 diagnosis tool using chest X-Ray images, that would additionally differentiate between controls, pneumonia or COVID-19 groups. 
The paper describes the process followed to train a Convolutional Neural Network with a dataset of more than $79,500$ X-Ray images compiled from different sources, including more than $8,500$ COVID-19 examples. For the sake of evaluation and comparison of the models developed, three different experiments were carried out following three preprocessing schemes. 
The aim is to evaluate how preprocessing the data affects the results and improves its explainability.
Likewise, a critical analysis is carried out about different variability issues that might compromise the system and the effects on the performance. 
With the employed methodology, a $91.5\%$ classification accuracy is obtained, with a $87.4\%$ average recall for the worst but most explainable experiment, which requires a previous automatic segmentation of the lungs region. 

\end{abstract}

\begin{IEEEkeywords}
COVID-19, Deep Learning, Pneumonia, radiological imaging, chest X-Ray.
\end{IEEEkeywords}


\section{Introduction}
\label{sec:introduction}

\IEEEPARstart{T}{he} COVID-19 pandemic has rapidly become into one of the biggest health world challenges in recent years. 
The disease spreads at a fast pace: the reproduction number of COVID-19 ranged from $2.24$ to $3.58$ during the first months of the pandemic \cite{zhao2020preliminary}, meaning that, on average, an infected person transmitted the disease to $2$ or more people. 
As a result, the number of COVID-19 infections dramatically increased from just a hundred cases in January --almost all of them concentrated in China-- to more than $43$ million in November spread all around the world \cite{ECDC:2020}.

COVID-19 is caused by the coronavirus SARS-COV2, a virus that belongs to the same family of other respiratory disorders such as the \textit{Severe Acute Respiratory Syndrome} (SARS) and \textit{Middle East Respiratory Syndrome} (MERS).
The symptomatology of COVID-19 is diverse and arise after incubation of around $5.2$ days.
These might include fever, dry cough, and fatigue; although, headache, haemoptysis, diarrhoea, dyspnoea, and lymphopenia are also reported \cite{rothan2020epidemiology,chen2020epidemiological}.
In severe cases, an \textit{Acute Respiratory Distress Syndrome} (ARDS) might be developed by underlying pneumonia associated with the COVID-19.
For the most serious cases, the estimated period from the onset of the disease to death ranged from 6 to 41 days (with a median of 14 days), being dependent on the age of the patient and the status of the patient's immune system \cite{rothan2020epidemiology}.

\begin{figure*}[!hp]
\setlength{\tempheight}{0.18\textheight}
\settowidth{\tempwidth}{\includegraphics[height=\tempheight]{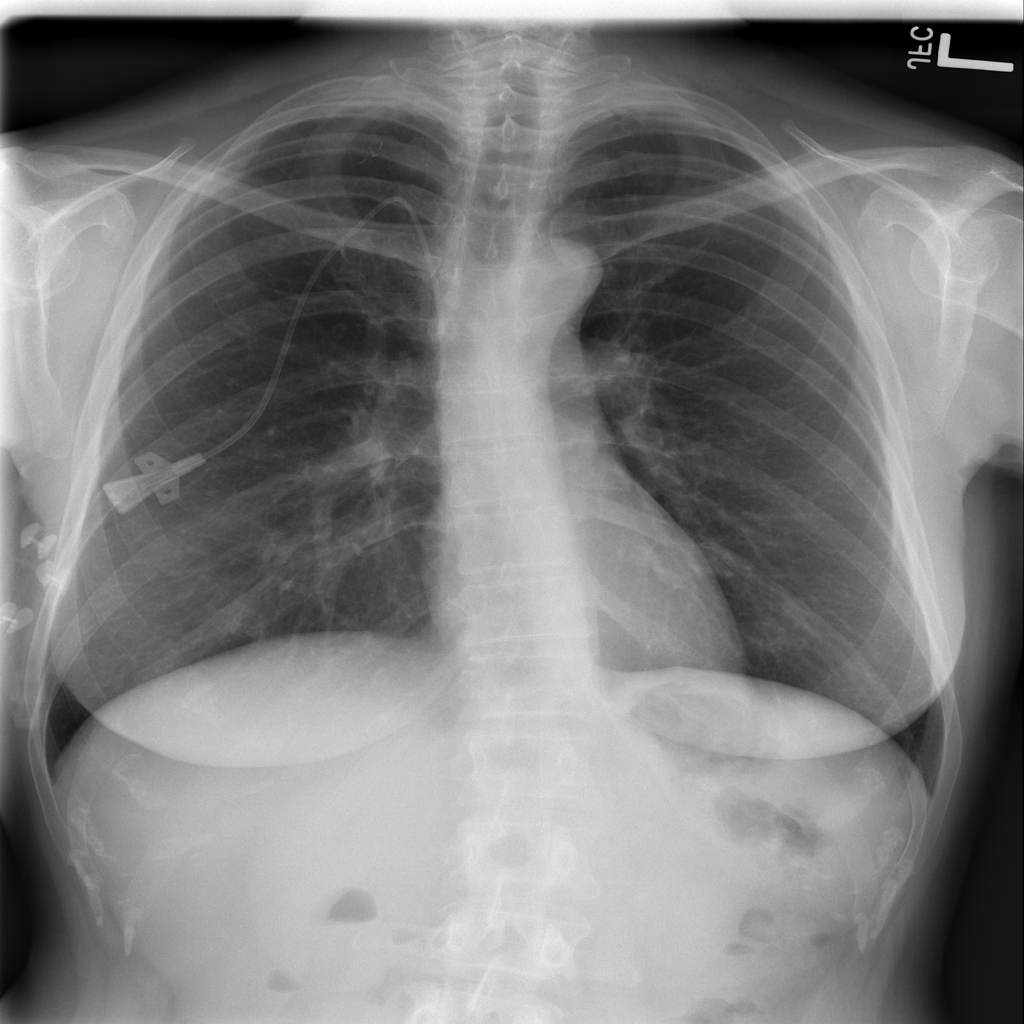}}
        \centering
        \hspace{0.1cm}
        \columnname{Control}\hspace{0.4cm}
        \columnname{Pneumonia}\hspace{0.3cm}
        \columnname{COVID-19}\\
        \rowname{Raw image}
        \subfloat[]
        {
            \label{rx-control}
            \includegraphics[width=0.25\textwidth]{images/CRXNIH__0__n__DX__PA__19169__5__00019169_017.png}
        }
        \subfloat[]
        {
            \label{rx-Pneumonia}
            \includegraphics[width=0.25\textwidth]{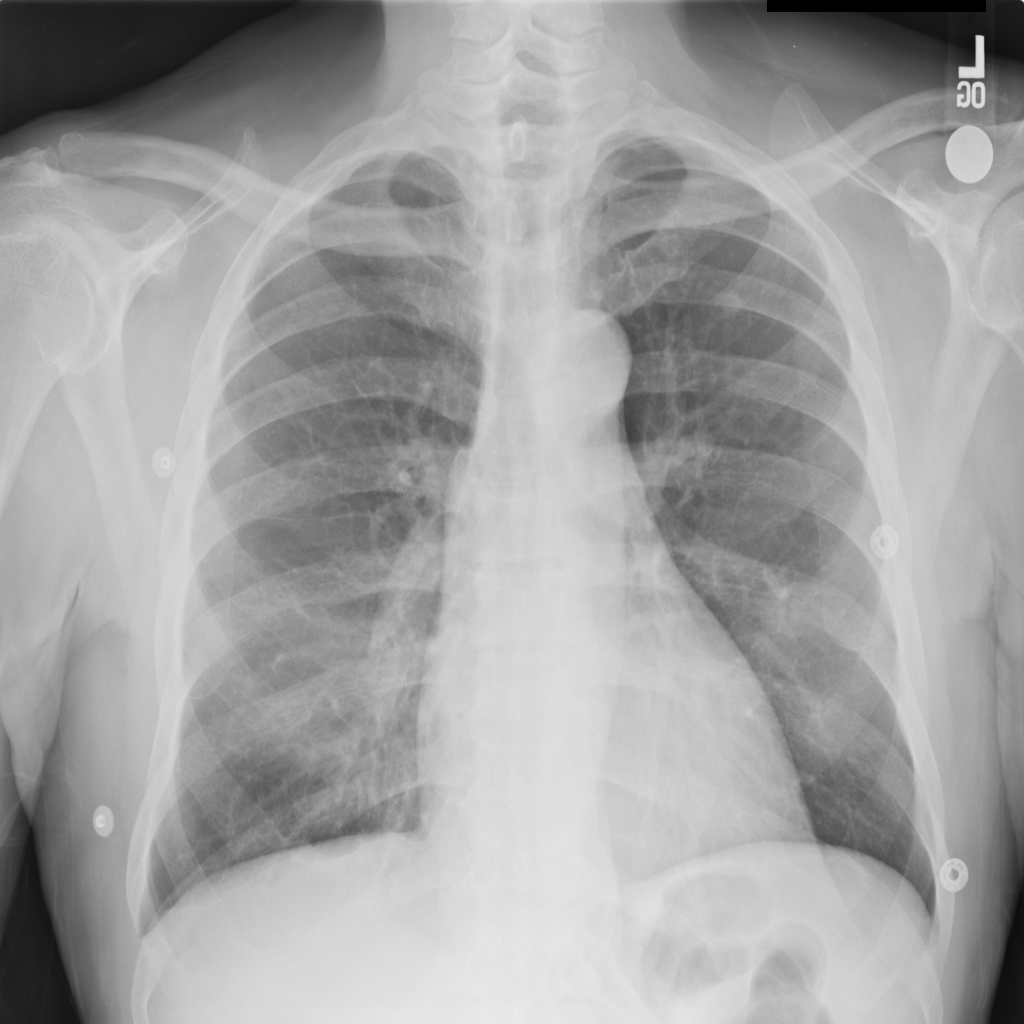}
        }
        \subfloat[]
        {
            \label{rx-COVID-19} 
            \includegraphics[width=0.25\textwidth]{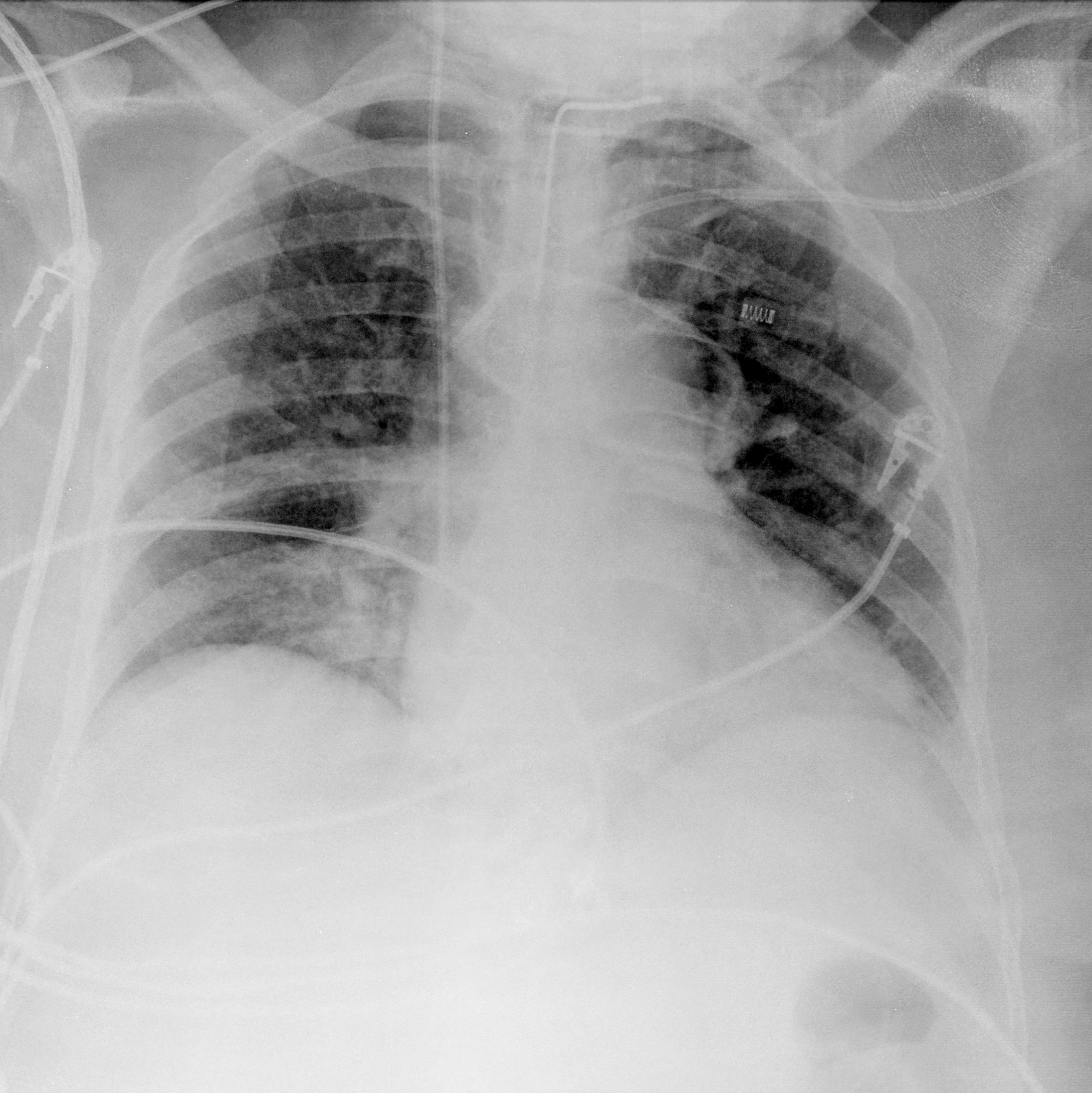}
        } 
        \\
        \rowname{Exp 1}
        \subfloat[]
        {
            \label{Exp1-control}
            \includegraphics[width=0.25\textwidth]{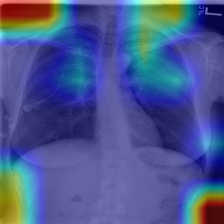}
        } 
            \subfloat[]
        {
            \label{Exp1-Pneumonia}  
            \includegraphics[width=0.25\textwidth]{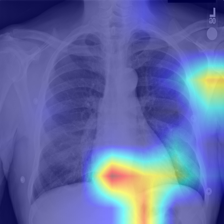}
        } 
        \subfloat[]
        {
            \label{Exp1-COVID-19}
            \includegraphics[width=0.25\textwidth]{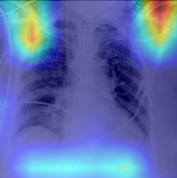}
        }
        \\
        \rowname{Exp 2}
        \subfloat[]
        {
            \label{Exp2-control}
            \includegraphics[width=0.25\textwidth]{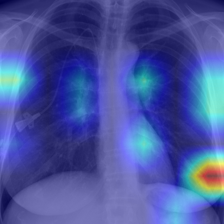}
        } 
            \subfloat[]
        {
            \label{Exp2-Pneumonia}  
            \includegraphics[width=0.25\textwidth]{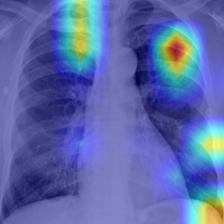}
        } 
        \subfloat[]
        {
            \label{Exp2-COVID-19}
            \includegraphics[width=0.25\textwidth]{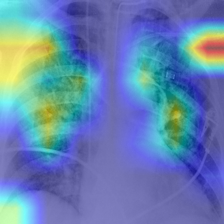}
        }
        \\ 
        \rowname{Exp 3}
        \subfloat[]
        {
            \label{Exp3-control}
            \includegraphics[width=0.25\textwidth]{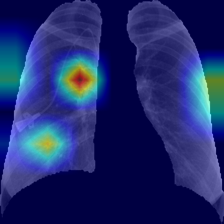}
        } 
        \subfloat[]
        {
            \label{Exp3-Pneumonia}  
            \includegraphics[width=0.25\textwidth]{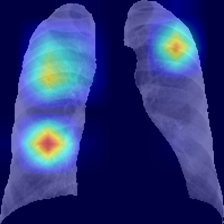}
        } 
        \subfloat[]
        {
            \label{Exp3-COVID-19}
            \includegraphics[width=0.25\textwidth]{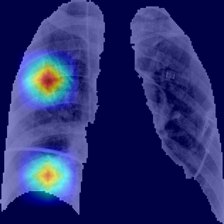}
        }
        
        \caption[]{ Experiments considered in the paper. \textbf{First row:} raw chest XR images belonging to the control, pneumonia, and COVID-19 classes. 
        \textbf{Second row:} Grad-CAM activation mapping for the XR images.
        Despite of the high accuracy of the methods, in some cases, the model focuses its attention in areas different from the lungs.
        \textbf{Third row:} Grad-CAM activation mapping after zooming in, cropping to a squared region of interest and resizing.
        Zooming to the region of interest forces the model to focus its attention to the lungs, but errors are still present.
        \textbf{Fourth row:} Grad-CAM activation mapping after a zooming and segmentation procedure.
        Zooming in and segmenting force the model to focus attention in the lungs. The black background represents the mask introduced by the segmentation procedure. 
        } 
        \label{fig:XR-examples}
\end{figure*}

Once the SARS-COV2 reaches the host at the lung, it gets into the cells through a protein called ACE2, which serves as the "opening" of the cell lock. 
After the genetic material of the virus has multiplied, the infected cell produces proteins that complement the viral structure to produce new viruses. 
Then, the virus destroys the infected cell, leave it and infect new cells. 
The destroyed cells produce radiological lesions \cite{pan2020initial,pan2020imaging,zhou2020coronavirus} such as consolidations and nodules in the lungs, that are observable in the form of ground-glass opacity regions in the XR images (Fig. \ref{rx-COVID-19}).
These lesions are more noticeable in patients assessed $5$ or more days after the onset of the disease, and especially in those older than $50$ \cite{song2020emerging}.
Findings also suggest that patients recovered from COVID-19 have developed pulmonary fibrosis \cite{Hosseiny2020}, in which the connective tissue of the lung gets inflamed. 
This leads to a pathological proliferation of the connective tissue between the alveoli and the surrounding blood vessels.
Given the aforementioned, radiological imaging techniques --using plain chest \textit{X-Ray} (XR) and/or thorax \textit{Computer Tomography} (CT)-- have become crucial diagnosis and evaluation tools to identify and assess the severity of the infection.

Since the declaration of the COVID-19 pandemic by the World Health Organization, four major key areas were identified to reduce the impact of the disease in the world: to prepare and be ready; detect, protect, and treat; reduce transmission; and innovate and learn
\cite{WHO_pandemic:2020}.
Concerning the area of detection, big efforts have been taken to improve the diagnostic procedures of COVID-19.
To date, the gold standard in the clinic is still a molecular diagnostic test based on a \textit{polymerase chain reaction} (PCR), which is precise but time-consuming, requires specialized personnel and laboratories and is in general limited by the capacities and resources of the health systems. 
This poses difficulties due to the rapid rate of growth of the disease. 
An alternative to PCR is the rapid tests such as those based in \textit{real-time reverse transcriptase-polymerase chain reaction} (RT-PCR), as they can be more rapidly deployed, decrease the load of the specialized laboratories, require less specialized personal and provide faster diagnosis compared to traditional PCR. 
Other tests, such as those based on antigens, are now available, but are mainly used for massive testings (i.e. for non clinical applications) due to a higher chance of missing an active infection. In contrast with RT-PCR, which detect the virus's genetic material, antigen tests identify specific proteins on the surface of the virus, requiring a higher viral load, which significantly shortens the period of sensitivity. 
In clinical practice, the RT-PCR test is usually complemented with a chest XR, in such a manner that the combined analysis reduces the significant number of false negatives and, at the same time, brings additional information about the extent and severity of the disease. 
In addition to that, thorax CT is also used as a second row method for evaluation. 
Although the evaluation with CT provides more accurate results in early stages and have been shown to have greater sensitivity and specificity \cite{ai2020correlation}, XR imaging has become the standard in the screening protocols, since it is fast, minimally-invasive, low-cost, and requires simpler logistics for its implementation. 

In the search of rapid, more objective, accurate and sensitive procedures, which could complement the diagnosis and assessment of the disorder, a trend of research has emerged to employ clinical features extracted from thorax CT or chest XR for automatic detection purposes. A potential benefit of studying the radiological images also comes from the potentiality of medical imaging to characterize pneumonic states even in asymptomatic population \cite{chan2020familial}, although more research is needed in this field as the lack of findings in infected patients is also reported \cite{li2020chest}.
The consolidation of such technology will permit a speedy and accurate diagnosis of COVID-19, decreasing the pressure on microbiological laboratories in charge of the PCR tests, and providing more objective means of assessing the severity of the disease.
To this end, techniques based on deep learning have been employed to characterize XR with promising results.
Although it would be desirable to employ CT for detection purposes, some major drawbacks are often present, including higher costs, a more time-consuming procedure, the necessity of thorough hygienic protocols not to spread infections, and the requirement of specialized equipment that might not be readily available in hospitals or health centres.
By contrast, XR is available as a first screening test in many hospitals or health centres, at lower expenses and with less time-consuming imaging procedures. 

Several approaches for COVID-19 detection based on chest XR images and different deep learning architectures have been published in the last few months, reporting classification accuracies around 90\% or higher. However, the central analysis in most of those works have focused on the variations of network architectures and less attention has been pay to the variability factors that a real solution should tackled before it can be deployed in the medical setting. In this sense, no analysis have been provided to demonstrate the reliability of the predictions made by the networks, which in the context of medical solutions acquires special relevance. Moreover, most of the works in the state of the art have validated their results with data sets containing dozens or a few hundreds of COVID-19 samples, limiting the impact of the proposed solutions.

With these antecedents in mind, this paper uses a deep learning algorithm based on CNN, data augmentation and regularization techniques to handle data imbalance, for the discrimination between COVID-19, controls and other types of pneumonia. The methods are tested with the largest corpus to date known by the authors. Three different sets of experiments were carried out in the search for the most suitable and coherent approach. To this end, the paper also uses explainability techniques to gain insight about the manners on how the neural network learns, and interpretability in terms of the overlaping among the regions of interest selected by the network and those that are more likely affected by COVID-19.
A critical analysis of factors that affect the performance of automatic systems based on deep learning is also carried out.

This paper is organized as follows: section \ref{background} presents some background and antecedents on the use of deep learning for COVID-19 detection. section \ref{sec:methodology} presents the methodology, section \ref{sec:results} presents the results obtained, whereas \ref{sec:disscon} presents the discussions and main conclusions of this paper. 

\section{Background}
\label{background}

A large body of research has emerged on the use of \textit{Artificial Intelligence} (AI) for the detection of different respiratory diseases using plain XR images. 
For instance, in \cite{rajpurkar2017chexnet} authors developed a $121$-layer \textit{Convolutional Neural Network} (CNN) architecture, called Chexnet, which was trained with a dataset of $100,000$ XR images for the detection of different types of pneumonia. The study reports an area under the \textit{Receiving Operatinng Characteristic}  (ROC) curve of $0.76$ in a multiclass scenario composed of $14$ classes.

Directly related to the COVID-19 detection, three CNN architectures (ResNet50, InceptionV3 and InceptionResNetV2) were considered in \cite{narin2020automatic}, using a database of just $50$ controls and $50$ COVID-19 patients. The best accuracy ($98\%$) was obtained with ResNet50.
In \cite{hemdan2020covidx}, seven different deep CNN models were tested using a corpus of $50$ controls and $25$ COVID-19 patients. The best resuts were attained with the VGG19 and DenseNet models, obtaining F1-scores of $0.89$ and $0.91$ for controls and patients.
The COVID-Net architecture was proposed in \cite{wang2020covid}. The net was trained with an open repository, called COVIDx, composed of $13,975$ XR images, although only $358$ --coming from $266$ patients-- belonged to the COVID-19 class. 
The attained accuracy was of $93.3\%$. 
In \cite{zhang2020covid} a deep anomaly detection algorithm was employed for the detection of COVID-19, in a corpus of $100$ COVID-19 images (taken from $70$ patients), and $1,431$ control images (taken from $1008$ patients). $96\%$ of sensitivity and $70\%$ of specificity was obtained.
In \cite{Islam2020}, a combination of a CNN for feature extraction and a \textit{Long Short Term Memory Network} (LSTM) for classification were used for automatic detection purposes. The model was trained with a corpus gathered from different sources, consisting of $4,575$ XR images: $1,525$ of COVID-19 (although $912$ come from a repository applying data augmentation), $1,525$ of pneumonia, and $1,525$ of controls. 
In a $5$-folds cross-validation scheme, a $99$\% accuracy was reported. 
In \cite{Civit-Masot2020}, the VGG16 network was used for classification, employing a database of $132$ COVID-19, $132$ controls and $132$ pneumonia images. 
Following a hold-out validation, about $100$\% accuracy was obtained identifying COVID-19, being lower on the other classes.

By using transfer-learning based on the Xception network, authors in \cite{NarayanDas2020} adapted a model for the classification of COVID-19. Experiments were carried out in a database of $127$ COVID-19, $500$ controls and $500$ patients with pneumonia gathered from different sources, attaining about $97$\% accuracy.
A similar approach, followed in \cite{Ozturk2020}, used the same corpus for the binary classification of COVID-19 and controls; and for the multi-class classification of COVID-19, controls and pneumonia. With a modification of the Darknet model for transfer-learning, and a $5$-folds cross-validation, a $98$\% accuracy in binary classification and $87$\% in multi-class classification was obtained.
Another Xception transfer-learning-based approach was presented in \cite{Khan2020}, but considering two multi-class classification tasks: i) controls vs. COVID-19 vs. viral pneumonia and bacterial pneumonia; ii) controls vs. COVID-19 vs. pneumonia.
To deal with the imbalance of the corpus, undersampling was used to randomly discard registers from the larger classes, obtaining $290$ COVID-19, $310$ controls, $330$ bacterial pneumonia and $327$ viral pneumonia chest XR images.
The reported accuracy in the $4$-class problem was of $89$\%, and of $94$\% in the $3$-class scenario. Moreover, in a $3$-class cross-database experiment, the accuracy was of $90$\%.
In \cite{Minaee2020}, four CNN networks (ResNet18, ResNet50, SqueezeNet, and DenseNet-121) were used for transfer learning. Experiments were performed on a database of $184$ COVID-19 and $5,000$ no-finding and pneumonia images. Reported results indicate a sensitivity of about $98$\% and a specificity of $93$\%.
In \cite{Apostolopoulos2020}, five state-of-the-art CNN systems --VGG19, MobileNetV2, Inception, Xception, InceptionResNetV2-- were tested on a transfer-learning setting to identify COVID-19 from control and pneumonia images.
Experiments were carried out in two partitions: one of $224$ COVID-19, $700$ bacterial pneumonia and $504$ control images; and another that considered the previous normal and COVID-19 data, but included $714$ cases of bacterial and viral pneumonia. 
The MobileNetV2 net attained the best results with $96$\% and $94$\% accuracy in the 2 and 3-classes classification, respectively.
In \cite{Apostolopoulos2020b}, the MobileNetV2 net was trained from scratch, and compared to one net based on transfer-learning and to another based on hybrid feature extraction with fine-tuning. Experiments performed in a dataset of $3905$ XR images of $6$ diseases indicated that training from the scratch outperforms the other approaches, attaining $87$\% accuracy in the multi-class classification and $99$\% in the detection of COVID-19. A system, also grounded on the InceptionNet and transfer-learning, was presented in \cite{Das2020}. Experiments were performed on $6$ partitions of XR images with COVID-19, pneumonia, tuberculosis and controls. 
Reported results indicate $99$\% accuracy, in a $10$-folds cross-validation scheme, in classification of COVID-19 from other classes.

In \cite{Togacar2020}, fuzzy colour techniques were used as a pre-processing stage to remove noise and enhance XR images in a 3-class classification setting (COVID-19, pneumonia and controls). The pre-processed images and the original ones were stacked. Then, two CNN models were used to extract features: MobileNetV2 and SqueezeNet. A feature selection technique based on social mimic optimization and a Support Vector Machine (SVM) were used.
Experiments were performed on a corpus of $295$ COVID-19, $65$ controls and $98$ pneumonia XR images, attaining about $99$\% accuracy.

Given the limited amount of COVID-19 images, some approaches have focused on generating artificial data to train better models.
In \cite{Waheed2020}, an auxiliary \textit{Generative Adversarial Network} (GAN) was used to produce artificial COVID-19 XR images from a database of $403$ COVID-19 and $1,124$ controls. 
Results indicated that data augmentation increased accuracy from $85$\% to $95$\% on the VGG16 net.
Similarly, in \cite{Loey2020}, GAN was used to augment a database of $307$ images belonging to four classes: controls, COVID-19, bacterial and viral pneumonia. Different CNN models were tested in a transfer-learning-based setting, including Alexnet, Googlenet, and Restnet18. The best results were obtained with Googlenet, achieving $99$\% in a multi-class classification approach.
In \cite{Toraman2020}, a CNN based on capsule networks (CapsNet), was used for binary (COVID-19 vs. controls) and multi-class classification (COVID-19 vs. pneumonia vs. controls). Experiments were performed on a dataset of $231$ COVID-19, $1,050$ pneumonia and $1,050$ controls XR images. Data augmentation was used to increase the number of COVID-19 images to $1,050$. On a $10$-folds cross-validation scheme, $97$\% accuracy for binary classification, and $84$\% multi-class classification were achieved.
The CovXNet architecture, based on depth-wise dilated convolution networks, was proposed in \cite{Mahmud2020}.
In a first stage, pneumonia (viral and bacterial) and control images were employed for pretraining. Then, a 
a refined model of COVID-19 is obtained using transfer learning.
In experiments using two-databases, $97$\% accuracy was achieved for COVID-19 vs. controls, and of $90$\% for COVID-19 vs. controls vs. bacterial and viral cases of pneumonia.
In \cite{Oh2020}, an easy-to-train neural network with a limited number of training parameters was presented. To this end, patch phenomena found on XR images were studied (bilateral involvement, peripheral distribution and ground-glass opacification) to develop a lung segmentation and a patch-based neural network that distinguished COVID-19 from controls. The basis of the system was the ResNet18 network. Saliency maps were also used to produce interpretable results. In experiments performed on a database of controls ($191$), bacterial pneumonia ($54$), tuberculosis ($57$) and viral pneumonia ($20$), about $89$\% accuracy was obtained. Likewise, interpretable results were reported in terms of large correlations between the activation zones of the saliency maps and the radiological findings found in the XR images. In addition to that, authors indicate that  when the lung segmentation approach was not considered the system's accuracy decreased to about $80$\%.
In \cite{Altan2020}, 2D curvelets transformations were used to extract features from XR images. A feature selection algorithm based on meta-heuristic was used to find the most relevant characteristics, while a CNN model based on EfficientNet-B0 was used for classification. Experiments were carried out in a database of $1,341$ controls, $219$ COVID-19, and $1,345$ viral pneumonia images, and $99$\% classification accuracy was achieved with the proposed approach. 
Multi-class and hierarchical classification of different types of diseases producing pneumonia (with $7$ labels and $14$ label paths), including COVID-19, were explored in \cite{Pereira2020}. 
Since the database of $1,144$ XR images was heavily imbalanced, different resampling techniques were considered. By following a transfer-learning approach based on a CNN architecture to extract features, and a hold-out validation with $5$ different classification techniques, a macro-avg F1-Score of $0.65$ and an F1-Score of $0.89$ were obtained for the multi-class and hierarchical classification scenarios respectively.
In \cite{Brunese2020}, a three-phases approach is presented: i) to detect the presence of pneumonia; ii) to classify between COVID-19 and pneumonia; and, iii) to highlight regions of interest of XR images. The proposed system utilized a database of $250$ images of COVID-19 patients, $2,753$ with other pulmonary diseases and $3,520$ controls. By using a transfer-learning system based on VGG16, about $0.97$ accuracy was reported.
A CNN-hierarchical approach using decision trees (based on ResNet18) was presented in \cite{Yoo2020}, on which a first tree classified XR images into the normal or pathological classes; the second identified tuberculosis; and the third COVID-19. Experiments were carried out on $3$ partitions obtained after having gathered images from different sources and data augmentation. The accuracy for each one of the decision trees --starting from the first-- was about $98$\%, $80$\%, and $95$\% respectively.

\subsection*{Issues affecting results in the literature}

Table \ref{tab:SoASummary} presents a summary of the state of the art in the automatic detection of COVID-19 based on XR images and deep learning. 
Despite the excellent results reported, the review reveals that some of the proposed systems suffer from certain shortcomings that affect the conclusions that can be extracted from them, limiting the possibility to be transferred to the clinical environment. Likewise, there exist variability factors that have not been deeply studied in these papers and which can be regarded as important.

For instance, one of the issues that affect most the reviewed systems to detect COVID-19 from plain chest XR images is the use of very limited datasets, which compromises their generalization capabilities.

Indeed, from the authors' knowledge, to date, the paper employing the largest database of COVID-19 considers $1,525$ XR images gathered from different sources. However, from these, $912$ belong to a data augmented repository, which does not include additional information about the initial number of files or the number of augmented images.
In general terms, most of the works employ less than $300$ COVID-19 XR images, having systems that use as few as $50$ images. This is however understandable since some of these works were published at the onset of the pandemics when the number of available registers was limited. 

On the other hand, a good balance in the age of the patients is considered important to avoid the model learn age-specific features. However, several previous works have used XR images from children to populate the pneumonia class \footnote{First efforts used the RSNA Pneumonia Detection Challenge dataset, which is focused on the detection of pneumonia cases in children.  https://www.kaggle.com/c/rsna-pneumonia-detection-challenge/overview}. This might be biasing the results given the age differences with respect to COVID-19 patients.

Despite many works in the literature report a good performance in the detection of COVID-19, most of the approaches follow a brute force approach exploiting the potentiality of deep learning to correlate with the outputs (i.e. the class labels), but providing low interpretability and explainability of the process. 
It means that it is unclear if the good results are due to the actual capability of the system to extract information related to the pathology or due to the capabilities of the system to learn other aspects biasing and compromising the results. 
As a matter of example, just one of the works reported in the literature follows a strategy that forces the network to focus in the most significant areas of interest for COVID-19 detection \cite{Oh2020}. 
It does so, by proposing a methodology based on a semantic segmentation of the lungs.
In the remaining cases, it is unclear if the models are analyzing the lungs, or if they are categorizing given any other information available, which might be interesting for classification purposes but might lack diagnostic interest.
This is relevant, as in all the analyzed works in the literature, pneumonia and controls classes come from a certain repository, whereas others such as COVID-19 comes from a combination of sources and repositories.
Having classes with different recording conditions might certainly affect the results, and as such, a critical study about this aspect is needed.
In the same line, other variability issues such as the sensor technology that is employed, the type of projection used, the sex of the patients, and even age, require a thorough study.

Finally, the review revealed that most of the published papers showed excellent correlation with the disease but low interpretability and explainability (see Table \ref{tab:SoASummary}).
Indeed, in clinical practice, it is often more desirable to obtain interpretable results that correlate to pathological conditions, or to a certain demographic or physiological variable, than a black box system that simply states a binary or a multiclass decision.
From the revision of literature, only \cite{Oh2020} and \cite{Mahmud2020} partially addressed this aspect. Thus, further research on this topic is needed.

With these ideas in mind, this paper addresses these aspects by training and testing with a wide corpus of RX images, proposing and comparing two strategies to preprocess the images, analyzing the effect of some variability factors, and providing some insights towards more explainable and interpretable results. 
The major goal is presenting a critical overview of these aspects since they might be affecting the modelling capabilities of the deep learning systems for the detection of COVID-19. 

\begin{table*}[htbp]
\caption{Summary of the literature in the field}
\begin{tabular}{@{}lp{5.5cm}lllllp{2cm}l@{}}
\toprule
\multirow{2}{*}{\textbf{Ref.}} & 
\multirow{2}{*}{\textbf{Architecture}}  & 
\multicolumn{3}{c}{\textbf{Number of cases}}  & 
\multirow{2}{*}{\textbf{Classes}} & \multirow{2}{*}{\textbf{\begin{tabular}[c]{@{}l@{}}Performance\\ metrics\end{tabular}}} & 
\multirow{2}{*}{\textbf{\begin{tabular}[c]{@{}l@{}}Lung\\segment.\end{tabular}}} & 
\multirow{2}{*}{\textbf{Explainable}} \\ \cmidrule(lr){3-5}
                               &                                                            & \textbf{COVID-19} & \textbf{Controls} & \textbf{Others} &                                   &                                                                                         &                                        &                                       \\ \midrule
\cite{narin2020automatic}      & InceptionV3, InceptionResNetV2, ResNet50                   & 50                  & 50                  &      --             & 2                                 & Acc=98\%                                                                                & N                                      & N                                     \\
\cite{hemdan2020covidx}        & VGG19, DenseNet                                            & 25                  & 50                  &      --             & 2                                 & AvF1=0.90                                                                               & N                                      & N                                     \\
\cite{wang2020covid}           & COVID-Net, ResNet50, VGG19                                                   & 358                  &    8066                 &      5538             & 3                                 & Acc=93.3\%                                                                                & N                                      & N                                     \\
\cite{zhang2020covid}          &  EfficientNet                                                          & 100                 & 1431                &     --              & 2                                 & Se=96\% Sp=70\%                                                                         & N                                      & N                                     \\
\cite{Islam2020}               & CNN + LSTM                                                 & 1525*                & 1525                & 1525             & 3                                 & Acc=99\%                                                                                & N                                      & N                                     \\
\cite{NarayanDas2020}          & Xception                                                   & 127                 & 500                 & 500               & 3                                 & Acc=97\%                                                                                & N                                      & N                                     \\
\cite{Ozturk2020}              & Darknet                                                    & 127                 & 500                 & 500               & 3                                 & Acc=87\%                                                                                & N                                      & N                                     \\
\cite{Khan2020}                & Xception                                                   & 290                 & 310                 & 657               & 3                                 & Acc=93\%                                                                                & N                                      & N                                     \\
\cite{Apostolopoulos2020}      & VGG19, MobileNetV2, Inception, Xception, InceptionResNetV2 & 224                 & 504                 & 700               & 3                                 & Acc=94\%                                                                                & N                                      & N                                     \\
\cite{Togacar2020}             & MobileNetV2, SqueezeNet                                    & 295                 & 65                  & 98                & 3                                 & Acc=99\%                                                                                & N                                      & N                                     \\
\cite{Das2020}                 & Inception                                       &  162                   &     2003                &   4650         & 3                                 & Acc=99\%                                                                                & N                                      & N                                     \\
\cite{Pereira2020}             &    Inception-V3                                                        &       90              &     1000                &      687            & 7                                 & AvF1=0.65                                                                                 & N                                      & N                                     \\
\cite{Waheed2020}              & VGG16                                                      & 403                 & 1124                &       --            & 2                                 & Acc=95\%                                                                                & N                                      & N                                     \\
\cite{Loey2020}                & Alexnet, Googlenet, Restnet18                              &  69                   &   79                  & 158                  & 4                                 & Acc=99\%                                                                                & N                                      & N                                     \\
\cite{Toraman2020}             &  Capsnet                                                          & 231                 & 1050                & 1050              & 3                                 & Acc=84\%                                                                                & N                                      & N                                     \\
\cite{Mahmud2020}              & CovXNet                                                    &     305                &      305               &    610               & 4                                 & Acc=90.2\%                                                                                & N                                      & Y                                     \\
\cite{Oh2020}                  & ResNet18                                                   &    180                 & 191                 & 131               & 4                                 & Acc=89\%                                                                                & Y                                      & Y                                     \\
\cite{Civit-Masot2020}         & VGG16                                                      & 132                 & 132                 & 132               & 3                                 &   AvF1=0.85                                                                                      & N                                      & N                                     \\
\cite{Yoo2020}                 & ResNet18                                                   &    162                 &      585               &         585          & 3                                 & Acc=95\%                                                                                & N                                      & N                                     \\
\cite{Minaee2020}              & ResNet18, ResNet50, SqueezeNet, DenseNet121               & 184                &    2400                &        2600           & 2                                 & Se=98\% Sp=92.9\%                                                                         & N                                      & N                                     \\
\cite{Brunese2020}             & VGG16                                                      & 250                 & 3520                & 2753              & 4                                 & Acc=97\%                                                                                & N                                      & N                                     \\
\cite{Altan2020}               & EfficientNet-B                                             & 219                 & 1341                & 1345              & 3                                 & Acc=99\%                                                                                & N                                      & N                                     \\ \bottomrule
\end{tabular}
* 912 coming from a repository of data augmented images.
\label{tab:SoASummary}%
\end{table*}

\section{Methodology} 
\label{sec:methodology} 
The design methodology is presented in the following section. The procedure that is followed to train the neural network is described firstly, along with the process that was followed to create the dataset. 
The network and the source code to train it are available at \url{https://github.com/jdariasl/COVIDNET}, so results can be readily reproduced by other researchers. 

\subsection{The network} 
\label{sec:network} 
The core of the system is a deep CNN based on the COVID-Net\footnote{Following the PyTorch implementation available at \url{https://github.com/IliasPap/COVIDNet}} proposed in \cite{wang2020covid}. 
Some modifications were made to include regularization components in the last two dense layers and a weighted categorical cross entropy loss function in order to compensate for the class imbalance.
The structure of the network was also refactored in order to allow gradient-based localization estimations \cite{Selvaraju_2019}, which are used after training in the search for an explainable model.

The network was trained with the corpus described in \ref{sec:corpus} using the Adam optimizer with a learning rate policy: the learning rate decreases when learning stagnates for a period of time (i.e., ’patience’). 
The following hyperparameters were used for training: learning rate=$2\textsuperscript{-5}$, number of epochs=$24$, batch size=$32$, factor=$0.5$, patience=$3$. 
Furthermore, data augmentation for pneumonia and COVID-19 classes was leveraged with the following augmentation types: horizontal flip, Gaussian noise with a variance of $0.015$, rotation, elastic deformation, and scaling. 
The variant of the COVID-Net was built and evaluated using the PyTorch library \cite{NEURIPS2019_9015}. 
The CNN features from each image are concatenated by a flatten operation and the resulting feature map is fed to three fully connected layers to generate a probability score for each class. The first two fully connected layers include dropout regularization of $0.3$ and ReLU activation functions. 
Dropout was necessary because the original network tended to overfit since the very beginning of the training phase. 

The input layer of the network rescales the images keeping the aspect ratio, with the shortest dimension scaled to $224$ pixels. Then, the input image is cropped to a square of $224 \times 224$ pixels located in the centre of the image. Images are normalized using a z-score function with parameters $mean=[0.485, 0.456, 0.406]$ and $std=[0.229, 0.224, 0.225]$, for each of the three RGB channels respectively. 
Even though we are working with grayscale images, the network architecture was designed to be pre-trained on a general purpose database including coloured images; this characteristic was kept in case it would be necessary to use some transfer learning strategy in the future.  

The output layer of the network provides a score for each of the three classes (i.e. control, pneumonia, or COVID-19), which is converted into three probability estimates --in the range $[0, 1]$-- using a softmax activation function. The final decision about the class membership is made according to the highest of the three probability estimates obtained.   

\subsection{The corpus}
\label{sec:corpus} 
The corpora used in the paper have been compiled from a set of \textit{Posterior-Anterior} (PA) and \textit{Anterior-Posterior} (AP) XR images from different public sources. 
The compilation contains images from participants without any observable pathology (controls or no findings), pneumonia, and COVID-19 cases. 
After the compilation, two subsets of images were generated, i.e. training and testing. Table \ref{tab:classDistribution} contains the number of images per subset and class. Overall, the corpus contains more than $70,000$ XR images, including more than $8,500$ images belonging to COVID-19 patients. 

\begin{table}[htbp]
  \centering
  \caption{Number of images per class for training and testing subsets}
    \begin{tabular}{@{}cccc@{}}
    \toprule
     {Subset} & \multicolumn{1}{l}{Control} & \multicolumn{1}{l}{Pneumonia} & \multicolumn{1}{l}{COVID-19} \\ \midrule
    Training & 45022 & 21707 & 7716 \\
    Testing & 4961 & 2407 & 857 \\ \bottomrule
    \end{tabular}%
  \label{tab:classDistribution}%
\end{table}%

The repositories of XR images employed to create the corpus used in this paper are presented next. 
Most of these contain solely registers of controls and pneumonia patients. Only the most recent repositories include samples of COVID-19 XR images. 
In all cases, the annotations were made by a specialist as indicated by the authors of the repositories.

The COVID-19 class is modelled compiling images coming from three open data collection initiatives: HM Hospitales COVID \cite{Hospitales2020}, BIMCV-COVID19 \cite{vay2020bimcv} and Actualmed COVID-19 \cite{Actualmed2020} chest XR datasets.
The final result of the compilation process is a subset of $8,573$ images from more than $3,600$ patients at different stages of the disease\footnote{Figures at the time the datasets were downloaded. The datasets are still open, and more data might be available in the next future}. 

Table \ref{tab:DemographicDistribution} summarizes the most significant characteristics of the datasets used to create the corpus, which are presented next: 

\subsubsection{HM Hospitales COVID-19 dataset}
This dataset was compiled by HM Hospitals \cite{Hospitales2020}. It contains all the available clinical information about anonymous patients with the SARS-CoV-2 virus who were treated in different centres belonging to this company since the beginning of the pandemic in Madrid, Spain. 

The corpus contains the anonymized records of $2,310$ patients, and collects different interactions in the COVID-19 treatment process including, among many other records, information on diagnoses, admissions, diagnostic imaging tests (RX and CT), treatments, laboratory results, discharge or death. The information is organized according to their content, all of them linked by a unique identifier. 
The dataset contains several radiological studies for each patient corresponding to different stages of the disease. Images are stored in a standard DICOM format. 
A total of $5,560$ RX images are available in the dataset, with an average of $2.4$ image studies per subject, often taken in intervals of two or more days. The histogram of the patients’ age is highly coherent with the demographics of COVID-19 in Spain (see Table \ref{tab:DemographicDistribution} for more details). 
Images were acquired with diverse devices, using different technologies, configurations, positions of the patient, and views. 
Only patients with at least one positive PCR test or positive immunological tests for SARS-CoV-2 were included in the study. 
The Data Science Commission and the Research Ethics Committee of HM Hospitales approved the current research study and the use of the data for this purpose. 

\subsubsection{BIMCV COVID19 dataset}
BIMCV COVID19 dataset \cite{vay2020bimcv} is a large dataset with chest radiological studies (XR and CT) of COVID-19 patients along with their pathologies, results of PCR and immunological tests, and radiological reports. It was recorded by the Valencian Region Medical Image Bank (BIMCV) in Spain. 
The dataset contains the anonymized studies of patients with at least one positive PCR test or positive immunological tests for SARS-CoV-2 in the period between February 26th and April 18th, 2020. Patients were identified by querying the Health Information Systems from 11 different hospitals in the Valencian Region, Spain. Studies were acquired using more than 20 different devices. 
The corpus is composed of a total of $3,013$ XR images, with an average of $1.9$ image studies per subject, taken in intervals of approximately two or more days. The histogram of the patients’ age is highly coherent with the demographics of COVID-19 in Spain (Table \ref{tab:DemographicDistribution}). All images are labelled with the technology of the sensor, but not all of them with the projection. Images were acquired with diverse devices, using different technologies, configurations, positions of the patient, and views. 
Only patients with at least one positive PCR test or positive immunological tests for SARS-Cov-2 were included in the study.

\begin{table*}[htbp]
  \centering
  \caption{Demographic data of the datasets used. Only those labels confirmed are reported}\label{tab:DemographicDistribution}%
    \begin{tabular}{@{}lllllllll@{}}
\toprule
              & Mean age $\pm$ std  & \# Males/\# Females & \# Images & AP/PA & DX/CR & COVID-19 & Pneumonia & Control \\ \midrule
HM Hospitales & 67,8 $\pm$ 15,7 & 3703/1857 *        & 5560   &  5018/542     &  1264/4296     &  Y   & N  & N  \\
BIMCV         & 62,4 $\pm$ 16,7 & 1527/1486 **       & 3013   & 1171/1217 & 1145/1868  &  Y   & N  & N  \\
ACT           & --                                    & --     & 188      & 30/155  & 126/59  &  Y   & N  & Y   \\
ChinaSet      & 35,4 $\pm$ 14,8 & 449/213             & 662    & 0/662    & 662/0   &  N   & Y  & Y  \\
Montgomery    & 51,9 $\pm$ 2,41 & 63/74               & 138    & 0/138      & 0/138 &  N   & Y  & Y   \\
CRX8        & 45,75 $\pm$ 16,83 & 34760/27030                  & 61790  & 21860/39930 & 61790/0   &  N   & Y  & Y   \\ 
CheXpert        & 62.38 $\pm$ 18,62 & 2697/1926                  & 4623  & 3432/1191 & --   &  N   & Y  & N   \\ 
MIMIC        & -- & --                  & 16399  & 10850/5549 & --   &  N   & Y  & N
\\\bottomrule
   & * 1377/929 patients  & ** 727/626 patients            &   &  &  &   &  &  
\end{tabular}
  
\end{table*}%

\subsubsection{Actualmed set (ACT)}

The actualmed COVID-19 Chest XR dataset initiative \cite{Actualmed2020} contains a series of XR images compiled by Actualmed and Universitat Jaume I (Spain). 
The dataset contains COVID-19 and control XR images, but no information is given about the place or date of recording and/or about the demographics. 
However, a metadata file is included. It contains an anonymized descriptor to distinguish among patients, and information about the XR modality, type of view and the class to which the image belongs. 

\subsubsection{China Set - The Shenzhen set}

The set was created by the National Library of Medicine, Maryland, USA in collaboration with the Shenzhen No.3 People’s Hospital at Guangdong Medical College in Shenzhen, China \cite{jaeger2014two}. 
The Chest XR images have been gathered from out-patient clinics and were captured as part of the daily routine using Philips DX Digital Diagnose systems .
The dataset contains normal and abnormal chest XR with manifestations of tuberculosis and includes associated radiologist readings.

\subsubsection{The Montgomery set}

This dataset was created by the National Library of Medicine in collaboration with the Department of Health and Human Services, Montgomery County, Maryland, USA.
It contains data from XR images collected under Montgomery County's tuberculosis screening program \cite{jaeger2014two, jaeger2013automatic}.
It contains a series of images of controls and tuberculosis patients, captured with a Eureka stationary X-ray machine (CR).
All images are de-identified and available in DICOM format. The set covers a wide range of abnormalities, including effusions and miliary patterns. 

\subsubsection{ChestX-ray8 dataset (CRX8)}

The ChestX-ray8 dataset \cite{wang2017chestx} contains $12,120$ images from $14$ common thorax disease categories from $30,805$ unique patients, compiled by the National Institute of Health (NIH). 
Natural language processing was used to extract the disease from the associated radiological reports. The labels are expected to be >90\% accurate and suitable for weakly-supervised learning.
For this study, the images labelled with 'no radiological findings' were used to be part of the control class, whereas the images annotated as 'pneumonia' were used for the pneumonia class.
In total $61,790$ images were used. 
Images annotated as pneumonia with other comorbidities were not included.


\subsubsection{CheXpert dataset}
CheXpert \cite{irvin2019chexpert} is a dataset of XR images created for an automated evaluation of medical imaging competitions, and contains chest XR examinations carried out in Stanford Hospital during $15$ years. For this study, we selected $4,623$ pneumonia images using those annotated as 'pneumonia' with and without another additional comorbidity. 
These comorbidities were never caused by COVID-19. The motivation to include pneumonia with comorbidities was to increase the number of pneumonia examples in the final compilation for this study, increasing the variability of this cluster.

\subsubsection{MIMIC-CXR Database}
MIMIC-CXR \cite{Johnson2019} is an open dataset complied from 2011 to 2016, and comprising de-identified chest RX from patients admitted to the Beth Israel Deaconess Medical Center. The dataset contains $371,920$ XR images associated with $227,943$ imaging studies. Each imaging study can pertain to one or more images, but most often are associated with two images: a frontal and a lateral view. The dataset is complemented with free-text radiology reports. 
In our study we employed the images for the pneumonia class. The labels were obtained from the agreement of the two methods indicated in \cite{Johnson2019}. The dataset reports no information about gender or age, thus, we assume that the demographics are similar to those of CheXpert dataset, and those of pneumonia \cite{Ramirez2017}.  

\subsection{Image Pre-processing} 

XR images were converted to uncompressed grayscale '.png' files, encoded with 16 bits, and preprocessed using the DICOM \textit{WindowCenter} and \textit{WindowWidth} details (when needed). All images were converted to a \textit{Monochrome 2} photometric interpretation. 
Initially, the images were not re-scaled, to avoid loss of resolution in later processing stages.

Only AP and PA views were selected. No differentiation was made between erect, either standing or sitting or decubitus. This information was inferred by a careful analysis of the DICOM tags, but also required a manual checking due to certain labelling errors.

\subsection{Experiments} 

The corpus collected from the aforementioned databases was processed to compile three different datasets of equal size to the initial one. Each of these datasets was used to run a different set of experiments. 

\subsubsection{Experiment 1. Raw data}

The first experiment was run using the raw data extracted from the different datasets. Each image is kept with the original aspect ratio. 
Only a histogram equalization was applied. 

\subsubsection{Experiment 2. Cropped image}

\begin{figure}[h!]
\centering
    \includegraphics[width=0.4\textwidth]{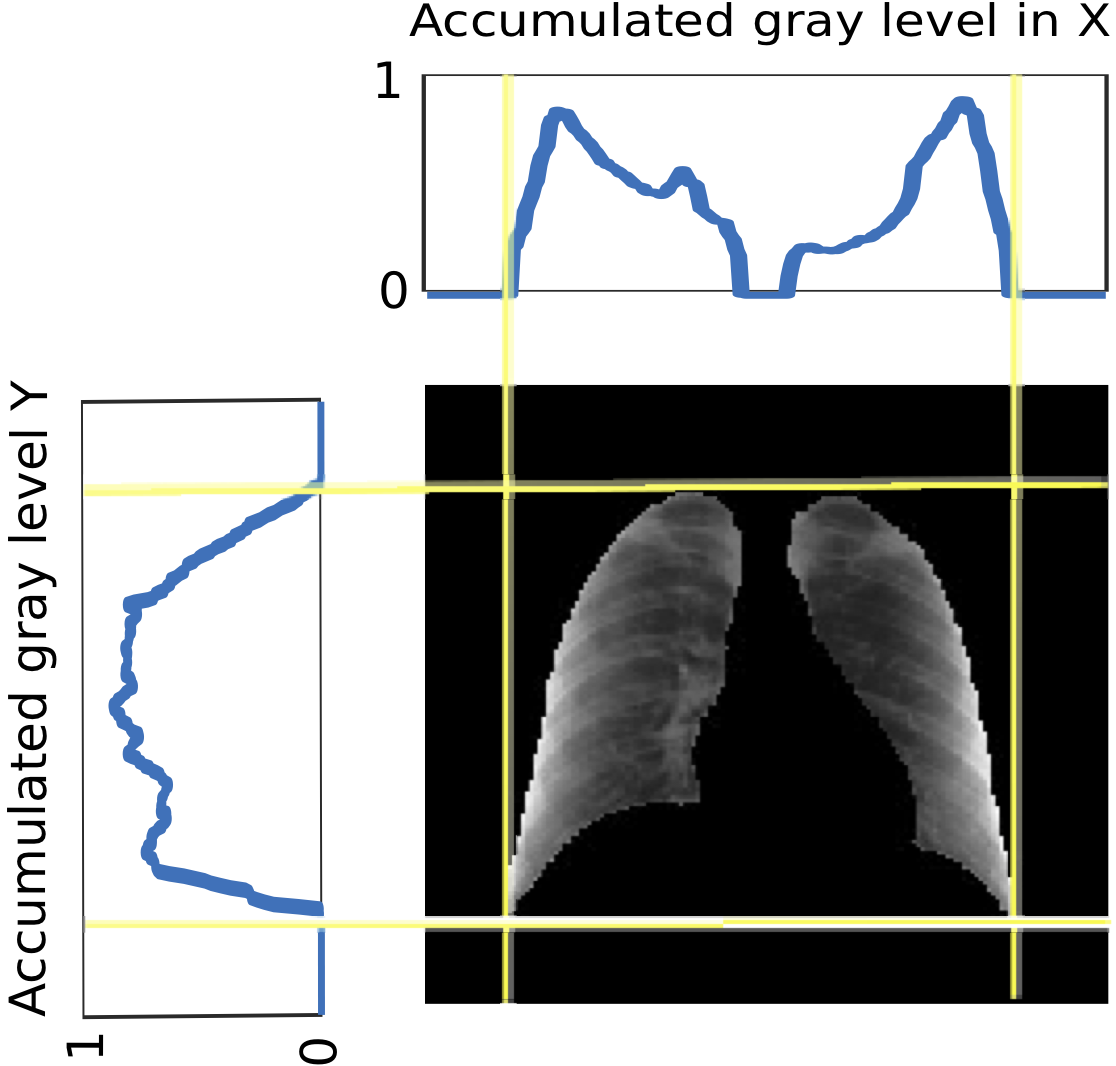}
    \caption{Identification of the squared region of interest. Plots in the top and left represent the normalized accumulated gray level in the vertical and horizontal dimension respectively. }
    \label{fig:Mascara}
\end{figure}

The second experiment consists of preprocessing the images by zooming in, cropping to a squared region of interest, and resizing to a squared image (aspect ratio $1:1$). The process is summarized in the following steps:

\begin{enumerate}
    \item Lungs are segmented from the original image using a U-Net semantic segmentation algorithm\footnote{Following the Keras implementation available at \url{https://github.com/imlab-uiip/lung-segmentation-2d}}. The algorithm used reports \textit{Intersection-Over-Union} (IoU) and Dice similarity coefficient scores of 0.971 and 0.985 respectively. 
    \item A black mask is extracted to identify the external boundaries of the lungs.
    \item The mask is used to create two sequences, adding the grey levels of the rows and columns respectively. These two sequences provide four boundary points, which define two segments of different lengths in the horizontal and vertical dimensions. 
    \item The sequences of added grey levels in the vertical and horizontal dimensions of the mask are used to identify a squared region of interest associated with the lungs, taking advantage of the higher added values outside the lungs (Fig. \ref{fig:Mascara}). The process to obtain the squared region requires identifying the middle point of each of the identified segments and cropping in both dimensions using the length of the longest of these two segments.
    \item The original image is cropped with a squared template placed in the centre of the matrix using the information obtained in the previous step. No mask is placed over the image.
    \item Histogram equalization of the image obtained. 
\end{enumerate}

This process is carried out to decrease the variability of the data, to make the training process of the network simpler, and to ensure that the region of significant interest is in the centre of the image with no areas cut. 

\subsubsection{Experiment 3. Lung segmentation}

The third experiment consists of preprocessing the images by masking, zooming in, cropping to a squared region of interest, and resizing to a squared image (aspect ratio $1:1$). The process is summarized in the following steps:

\begin{enumerate}
    \item Lungs are segmented from the original image using the same semantic segmentation algorithm used in experiment 2. 
    \item An external black mask is extracted to identify the external boundaries of the lungs.  
    \item The mask is used to create two sequences, adding the grey levels of the rows and columns respectively. 
    \item The sequences of added grey levels in the vertical and horizontal dimensions of the mask are used to identify a squared region of interest associated to the lungs, taking advantage of the higher added values outside them (Fig. \ref{fig:Mascara}).
    \item The original image is cropped with a squared template placed in the center of the image. 
    \item The mask is dilated with a $5 \times 5$ pixels kernel, and it is superimposed to the image. 
    \item Histogram equalization is applied only to the segmented area (i.e. the area corresponding to the lungs).
\end{enumerate}
 
 This preprocessing makes the training of the network much simpler and forces the network to focus the attention on the lungs region, removing external characteristics --like the sternum-- that might influence the obtained results.

\begin{table*}[!ht]
\centering
\caption{Performance measures for the three experiments considered in the paper}\label{tab:NumericResults}
\begin{tabular}{@{}lclccccc@{}}
\toprule
\multirow{2}{*}{\textbf{Experiment}} & \multirow{2}{*}{\textbf{Class}} & \multicolumn{6}{c}{\textbf{Measures}}                                                                                                                                \\ \cmidrule(l){3-8} 
                                     &                                 & \textbf{PPV}     & \textbf{Recall}    & \textbf{F1}      & \textbf{Acc}                      & \textbf{BAcc}                     & \textbf{GMR}             \\ \midrule
\multirow{3}{*}{\textbf{Exp. 1}}     & \textit{Pneumonia}              & 92.53 $\pm$ 1.13 & 94.20 $\pm$ 1.43   & 93.35 $\pm$ 0.68 & \multirow{3}{*}{91.67 $\pm$ 2.56} & \multirow{3}{*}{94.43 $\pm$ 1.36} & \multirow{3}{*}{93.00 $\pm$ 1.00} \\
                                     & \textit{Control}                & 93.35 $\pm$ 0.68 & 96.56 $\pm$   0.50 & 97.24 $\pm$ 0.23 &                                   &                                   &                                   \\
                                     & \textit{COVID-19}               & 91.67 $\pm$ 2.56 & 94.43 $\pm$  1.36  & 93.00 $\pm$ 1.00 &                                   &                                   &                                   \\ \midrule
\multirow{3}{*}{\textbf{Exp. 2}}     & \textit{Pneumonia}              & 84.02 $\pm$ 1.16 & 85.75 $\pm$ 1.46   & 84.86 $\pm$ 0.51 & \multirow{3}{*}{87.64 $\pm$ 0.74} & \multirow{3}{*}{81.35 $\pm$ 2.70} & \multirow{3}{*}{81.36 $\pm$ 0.42} \\
                                     & \textit{Control}                & 93.62 $\pm$ 0.76 & 92.67 $\pm$  0.69  & 93.14 $\pm$ 0.25 &                                   &                                   &                                   \\
                                     & \textit{COVID-19}               & 81.60 $\pm$ 3.33 & 81.35 $\pm$ 2.70   & 81.36 $\pm$ 0.42 &                                   &                                   &                                   \\ \midrule
\multirow{3}{*}{\textbf{Exp. 3}}     & \textit{Pneumonia}              & 85.26 $\pm$ 0.73 & 85.26 $\pm$ 0.73   & 87.42 $\pm$ 0.27 & \multirow{3}{*}{91.53 $\pm$ 0.20} & \multirow{3}{*}{87.64 $\pm$ 0.74} & \multirow{3}{*}{87.37 $\pm$ 0.84} \\
                                     & \textit{Control}                & 96.99 $\pm$ 0.17 & 94.48 $\pm$ 0.24   & 95.72 $\pm$ 0.15 &                                   &                                   &                                   \\
                                     & \textit{COVID-19}               & 78.52 $\pm$ 2.08 & 78.73 $\pm$ 2.80   & 78.57 $\pm$ 1.15 &                                   &                                   &                                   \\ \bottomrule
\end{tabular}

\end{table*}

 \begin{figure*}[!h]
    \setlength{\tempheight}{0.18\textheight}
    \settowidth{\tempwidth}{\includegraphics[height=\tempheight]{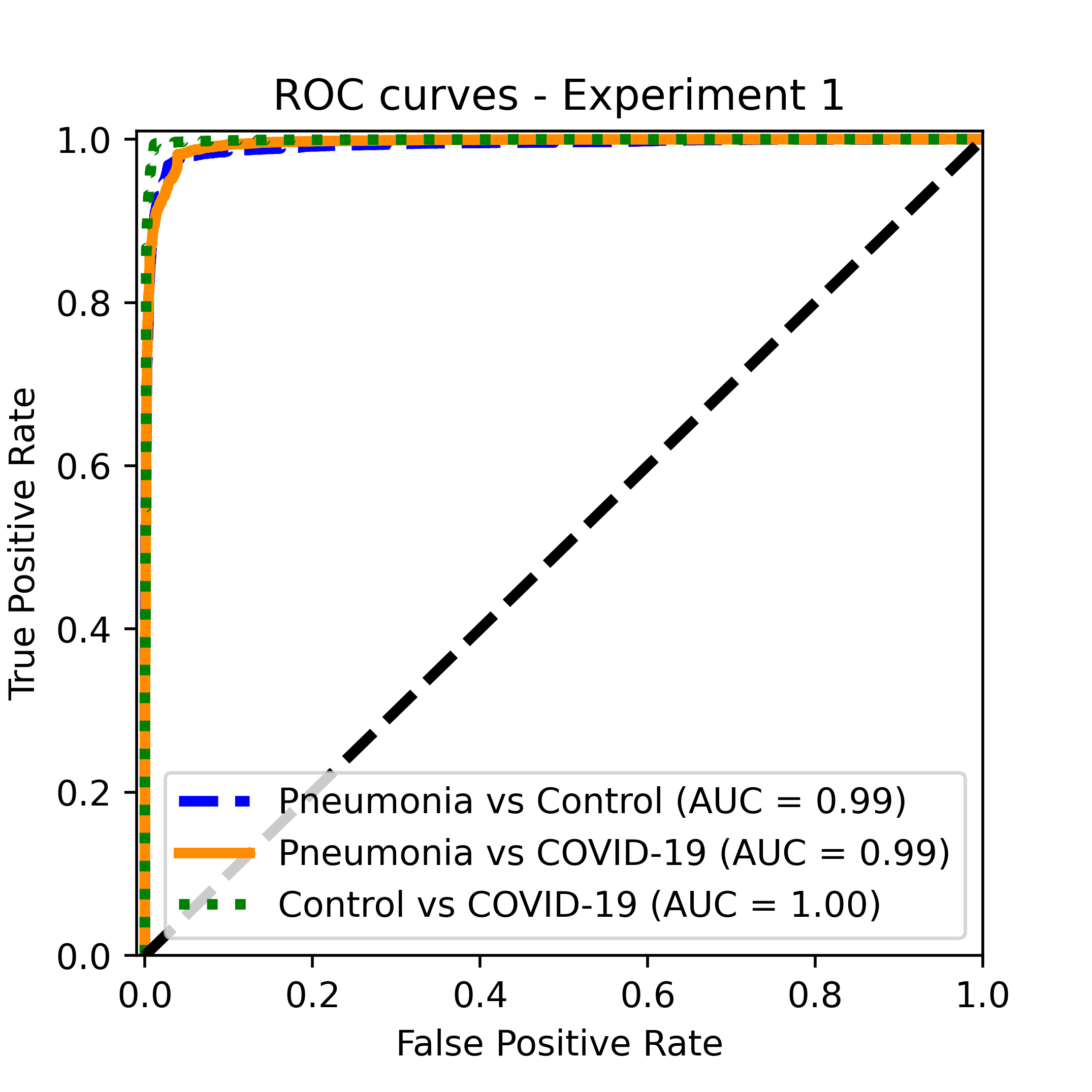}}
      \centering
      \hspace{\baselineskip}
      \columnname{Exp. 1}\hfil\vspace{-0.3cm}
      \columnname{Exp. 2}\hfil
      \columnname{Exp. 3}\hfil
      \subfloat[]
         {
             \label{ROC_original}
             \includegraphics[width=0.33\textwidth]{images/ROC_Original.png}
         }
         \subfloat[]
         {
             \label{ROC_Cropped}
             \includegraphics[width=0.33\textwidth]{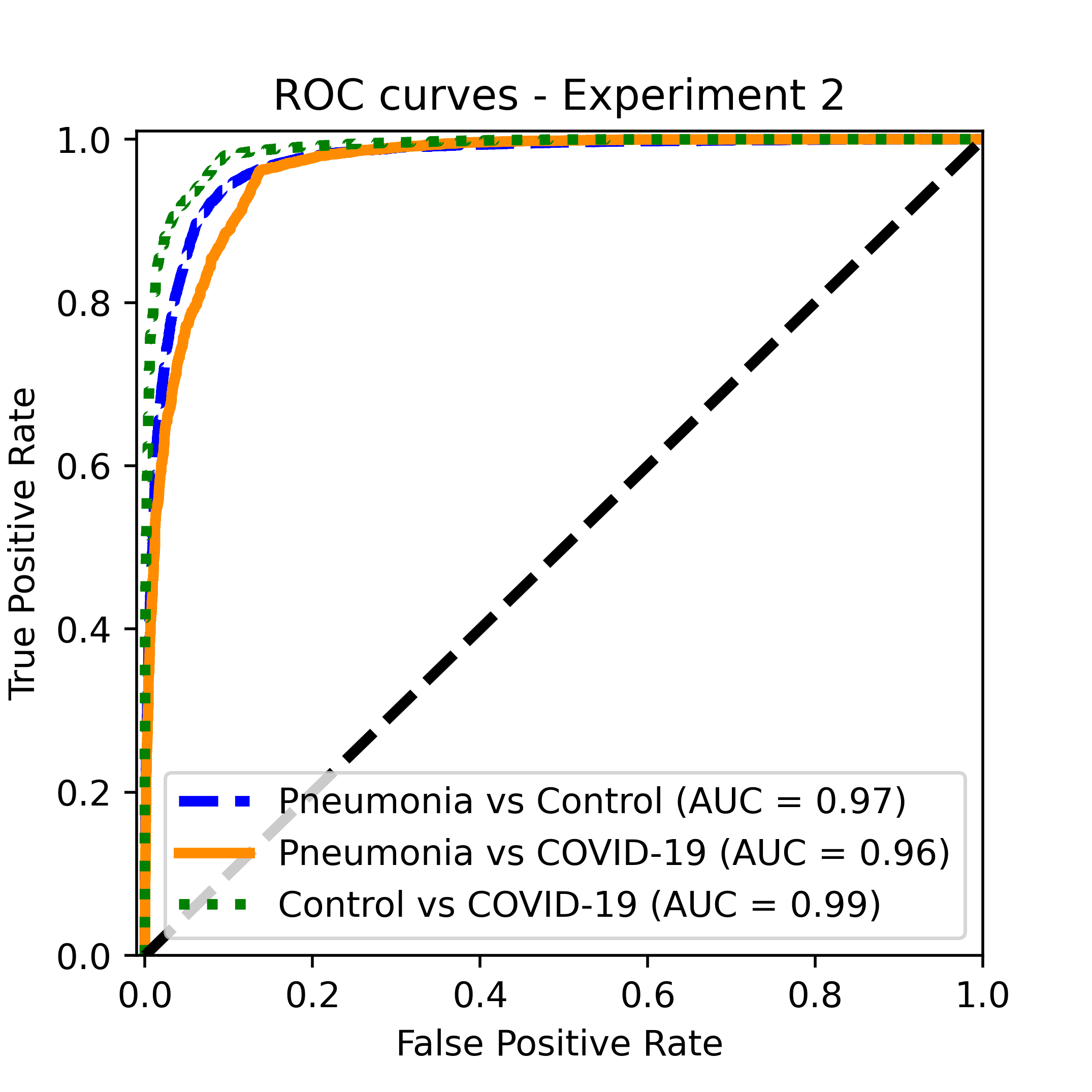}
         }
         \subfloat[]
         {
             \label{ROC_CropSeg} 
             \includegraphics[width=0.33\textwidth]{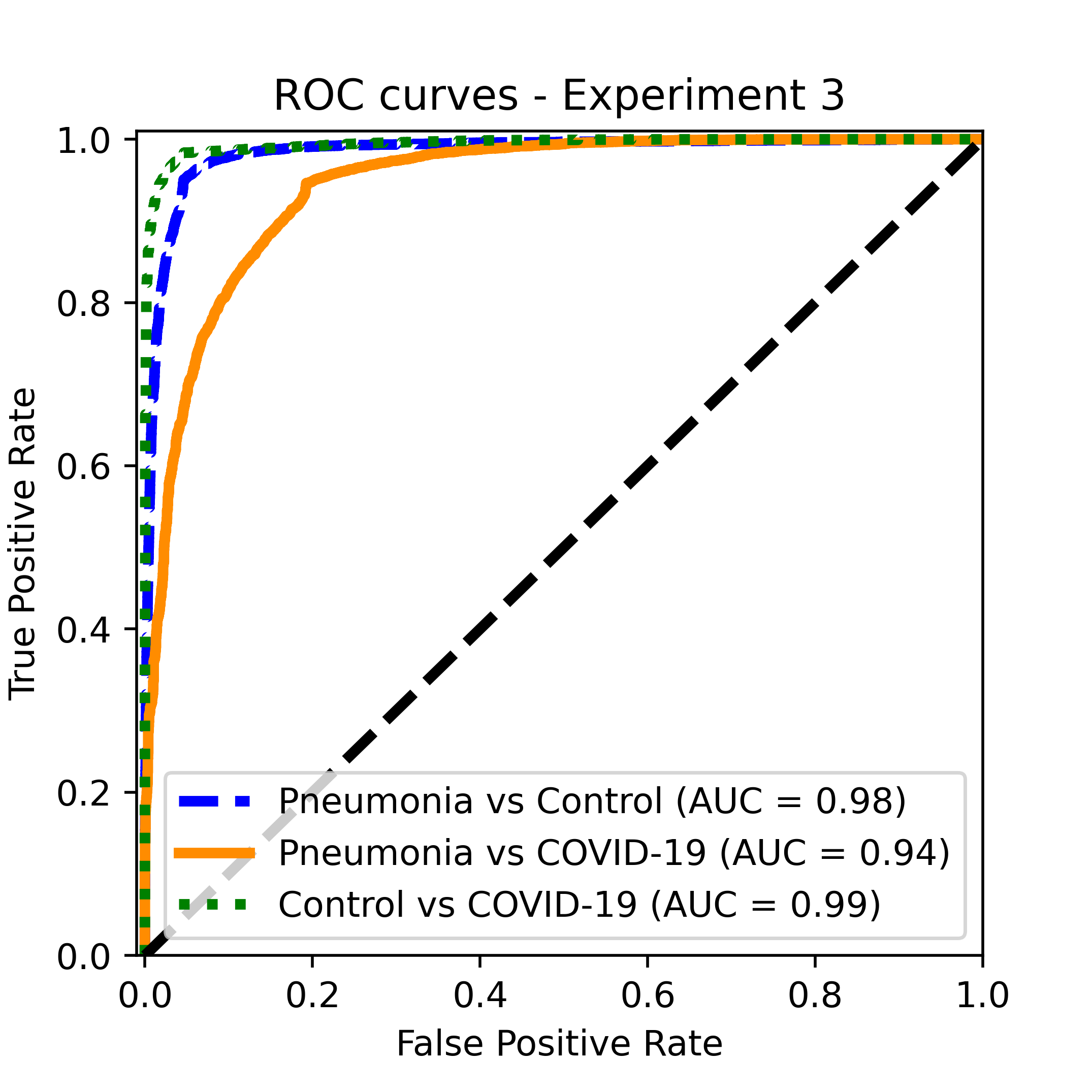}
         } \\ 
         \subfloat[]
         {
             \label{CM_original}
             \includegraphics[width=0.33\textwidth]{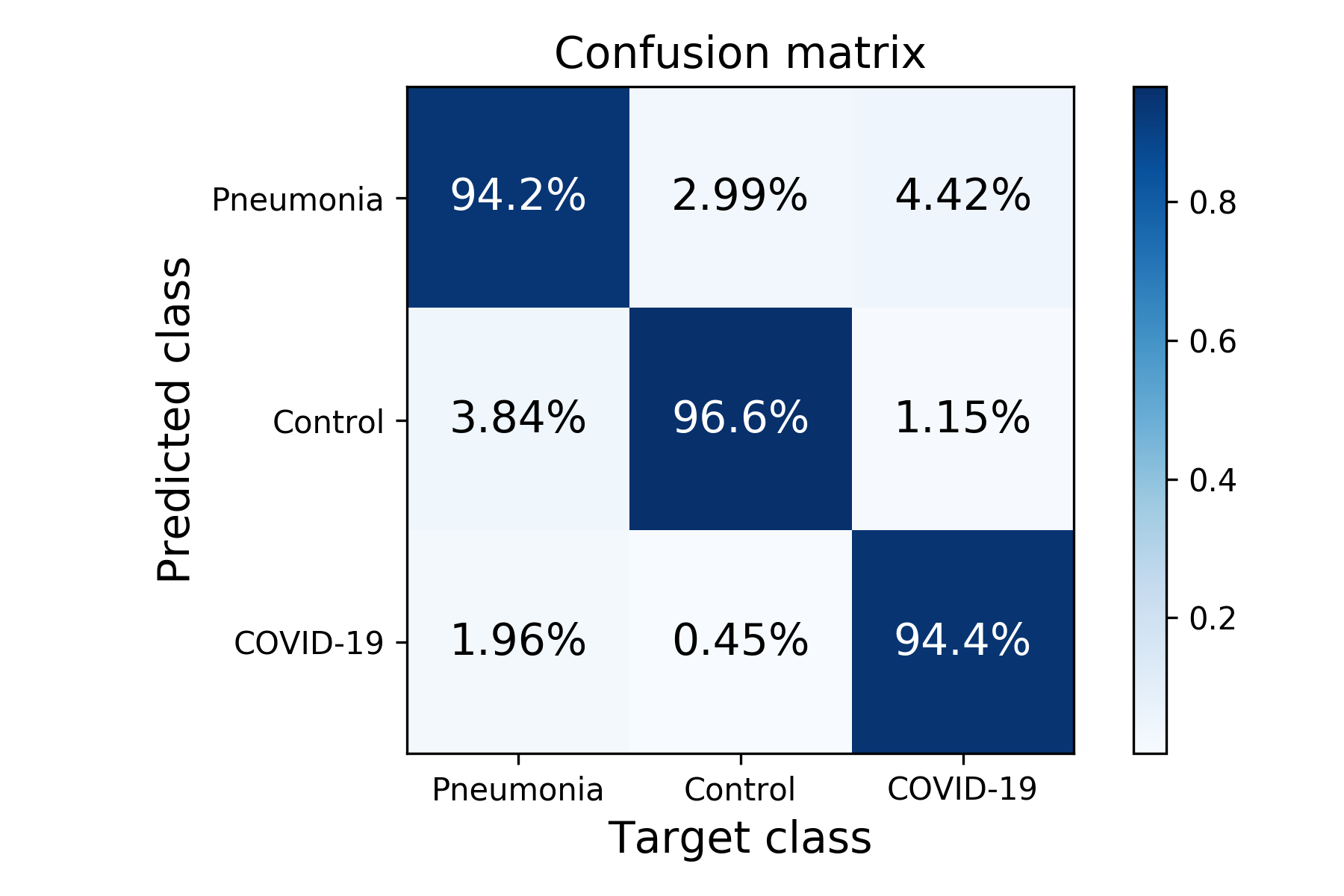}
         }
         \subfloat[]
         {
             \label{CM_Cropped}
             \includegraphics[width=0.33\textwidth]{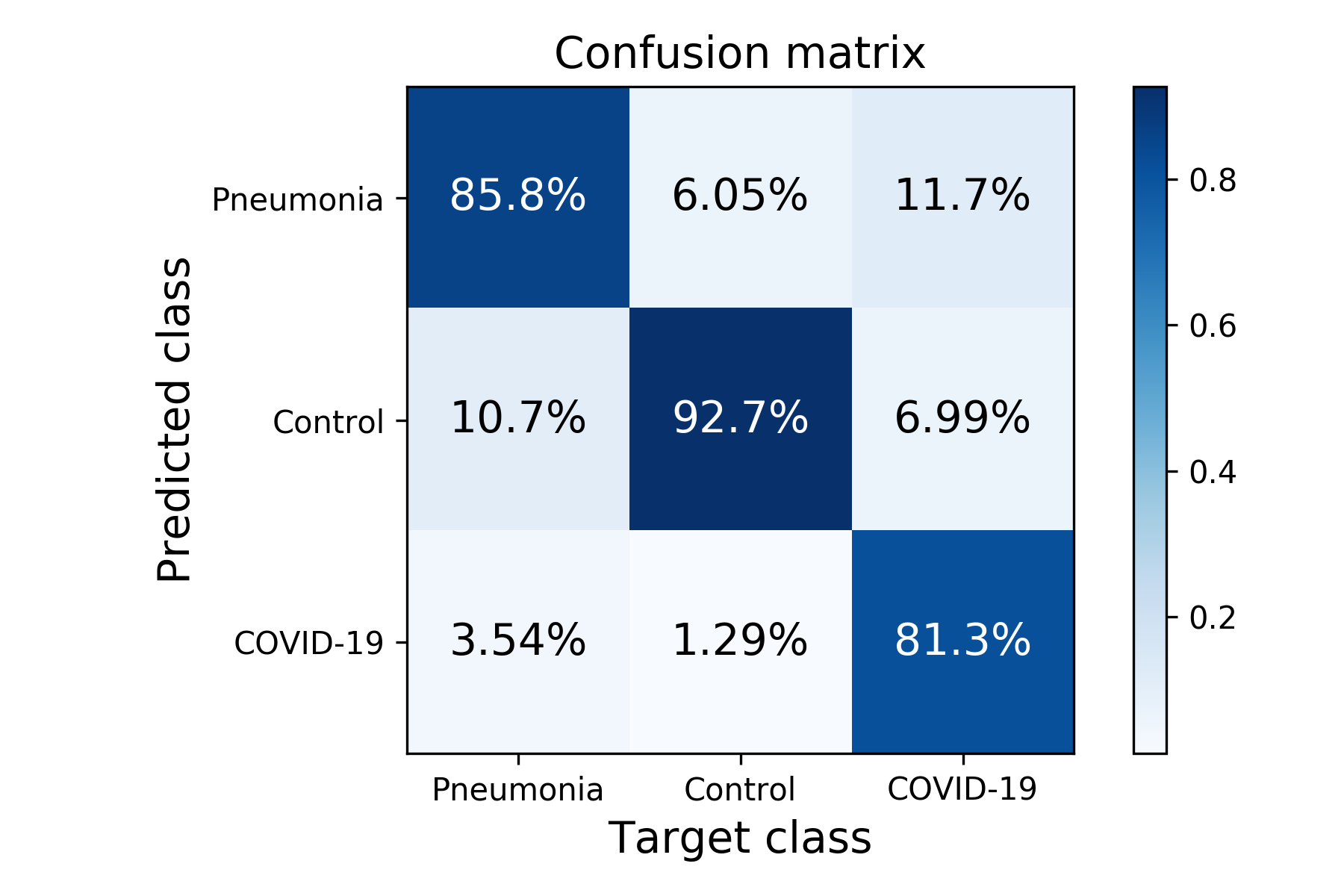}
         }
         \subfloat[]
         {
             \label{CM_CropSeg} 
             \includegraphics[width=0.33\textwidth]{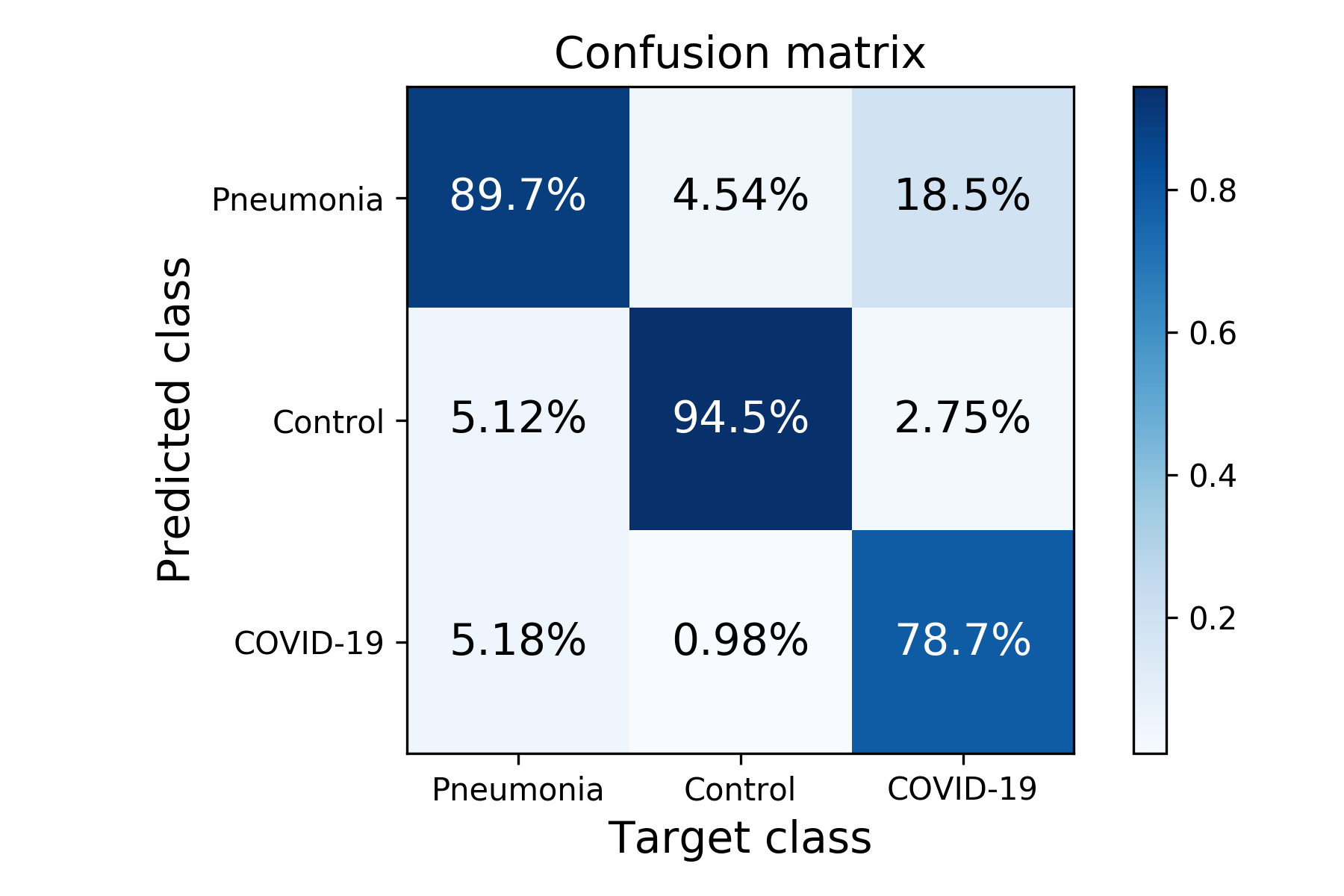}
         } 
         \caption{ROC curves and confusion matrices for each one of the experiments, considering each one of the classes separately. 
         \textbf{Top:} ROC curves. \textbf{Bottom:} Normalized confusion matrices. \textbf{Left:} Original images (experiment 1). \textbf{Center:} Cropped Images (experiment 2). \textbf{Right:} Segmented images (experiment 3). 
          } 
          \label{fig:ROC_CMatrix}
 \end{figure*}

\begin{figure}[h!]
\centering
    \includegraphics[width=0.4\textwidth]{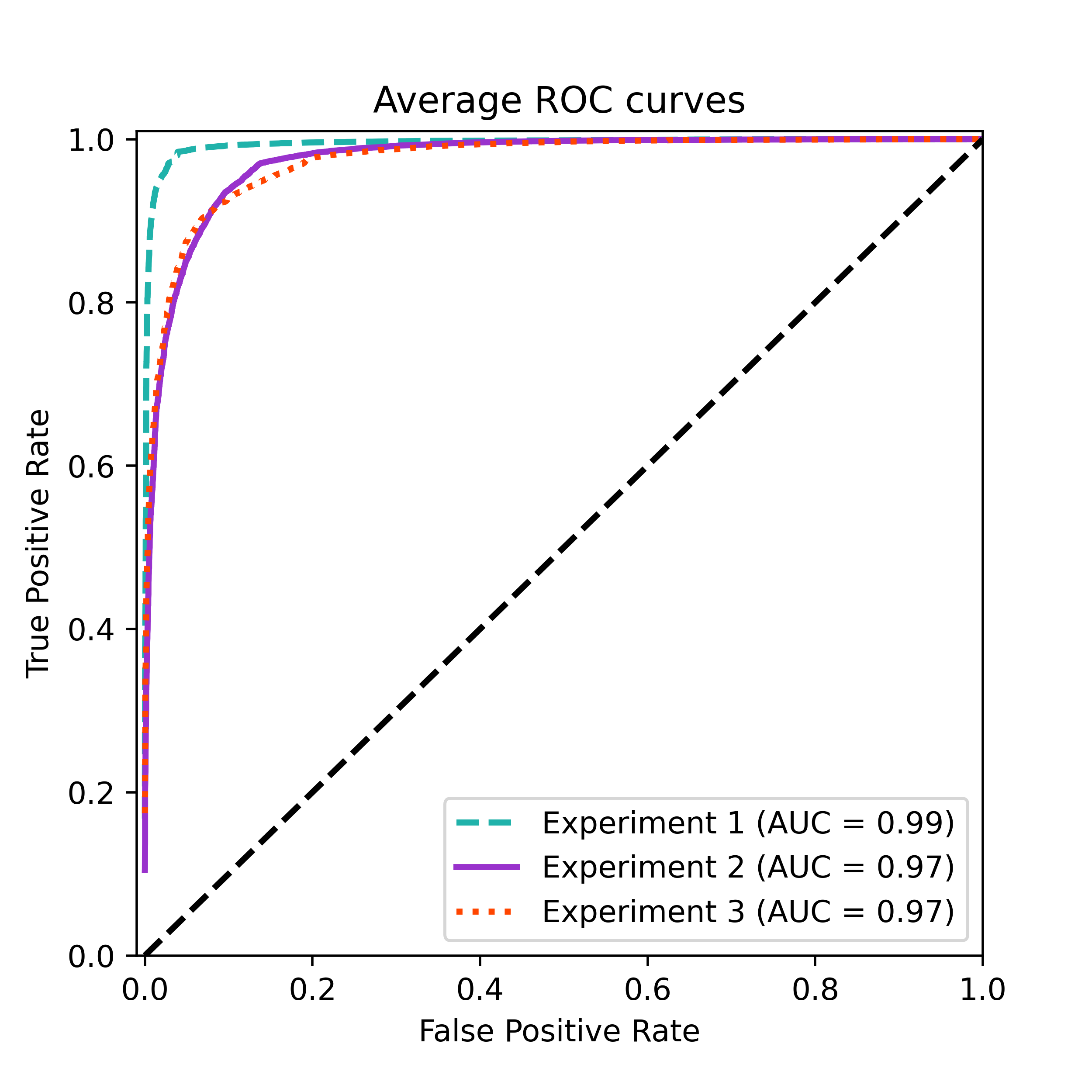}
    \caption{Average ROC curves for each experiment, including AUC values.}
    \label{fig:ROCThree}
\end{figure}

\subsection{Identification of the areas of significant interest for the classification} 

The areas of significant interest used by the CNN for discrimination purposes are identified using a qualitative analysis based on a \textit{Gradient-weighted Class Activation Mapping} (Grad-CAM) \cite{Selvaraju_2019}. 
This is an explainability method that serves to provide insights about the manners on how deep neural networks learn, pointing to the most significant areas of interest for decision-making purposes. 
The method uses the gradients of any target class to flow until the final convolutional layer, and to produce a coarse localization map which highlights the most important regions in the image identifying the class. 
The result of this method is a heat map like those presented in Fig. \ref{fig:XR-examples}, in which the colour encodes the importance of each pixel in differentiating among classes. 

\section{Results}
\label{sec:results} 

The model has been quantitatively evaluated computing the test \textit{Positive Predictive Value} (PPV), \textit{Recall}, \textit{F1-score} (F1), \textit{Accuracy} (Acc), \textit{Balanced Accuracy} (BAcc), \textit{Geometric Mean Recall} (GMR) and \textit{Area Under the ROC Curve} (AUC) for each of the three classes in the corpus previously described in section \ref{sec:corpus}.
The performance of the models is assessed using an independent testing set, which has not been used during development.
A $5$-folds cross-validation procedure has been used to evaluate the obtained results (Training/Test balance: 90/10 \%). 
The performance of the CNN network on the three experiments considered in this paper is summarized in Table \ref{tab:NumericResults}. 
Likewise, the ROC curves per class for each of the experiments, and the corresponding confusion matrices are presented in Fig. \ref{fig:ROC_CMatrix}. 
The global ROC curve displayed in Fig. \ref{fig:ROCThree} for each experiment summarizes the global performance of the experiments.

Considering experiment 1, and although slightly higher for controls, the detection performance remains almost similar for all classes (the PPV ranges from $91$-$93\%$) (Table \ref{tab:NumericResults}).
The remaining measures per class follow the same trend, with similar figures but better numbers for the controls.
ROC curves and confusion matrices of Fig. \ref{fig:ROC_CMatrix}a and Fig. \ref{fig:ROC_CMatrix}d point out that the largest source of confusion for COVID-19 is the pneumonia class. 
The ROC curves for each one of the classes reach in all cases AUC values larger than $0.99$, which, in principle is considered excellent.
In terms of global performance, the system achieves an Acc of $91\%$ and a BAcc of $94\%$ (Table \ref{tab:NumericResults}). This is also supported by the average ROC curve of Fig. \ref{fig:ROCThree}, which reveals the excellent performance of the network and the almost perfect behaviour of the ROC curve. Deviations are small for the three classes. 

When experiment 2 is considered, a decrease in the performance per class is observed in comparison to experiment 1. In this case, the PPV ranges from $81$-$93\%$ (Table \ref{tab:NumericResults}), with a similar trend for the remaining figures of merit.
ROC curves and confusion matrices in Fig. \ref{ROC_Cropped} and Fig. \ref{CM_Cropped} report AUC values in the range $0.96$-$0.99$, and an overlapping of the COVID-19 class mostly with pneumonia.
The global performance of the system -presented in the ROC curve of Fig. \ref{fig:ROCThree} and Table \ref{tab:NumericResults}- yields an AUC of $0.98$, an Acc of $87\%$ and a BAcc of $81\%$.

Finally, for the experiment 3, PPV ranges from $78\%-96\%$ (Table \ref{tab:NumericResults}). In this case, the results are slightly worse than those of experiment 2, with the COVID-19 class presenting the worse performance among all the tests. According to Fig. \ref{ROC_CropSeg}, AUCs range from $0.94$ to $0.98$.
Confusion matrix in Fig. \ref{CM_CropSeg} reports a large level of confusion in the COVID-19 class being labelled as pneumonia $18\%$ of the times. 
In terms of global performance the system reaches an Acc of $91\%$ and a BAcc of $87\%$ (Table \ref{tab:NumericResults}).
These results are consistent with the average AUC of $0.97$ shown in Fig. \ref{fig:ROCThree}.

\subsection{Explainability and interpretability of the models} 

The regions of interest identified by the network, were analyzed qualitatively using Grad-CAM activation maps \cite{Selvaraju_2019}. Results shown by the activation maps, permit the identification of the most significant areas in the image, highlighting the zones of interest that the network is using to discriminate.
In this regard, Fig. \ref{fig:XR-examples}, presents examples of the Grad-CAM of a control, a pneumonia, and a COVID-19 patient, for each of the three experiments considered in the paper. 
It is important to note that the activation maps are providing overall information about the behaviour of the network, pointing to the most significant areas of interest, but the whole image is supposed to be contributing to the classification process to a certain extent. 

The second row in Fig. \ref{fig:XR-examples} shows several prototipical results applying the Grad-CAM techniques to experiment 1. The examples show the areas of significant interest for a control, pneumonia and COVID-19 patient. 
The results suggest that the detection of pneumonia or COVID-19 is often carried out based on information that is outside the expected area of interest, i.e. the lung area. In the examples provided, the network focuses on the corners of the XR image or in areas around the diaphragm. 
In part, this is likely due to the metadata which is frequently stamped on the corners of the XR images.  
The Grad-CAM plots corresponding to the experiment 2 (third row of Fig \ref{fig:XR-examples}), indicates that the model still points towards areas which are different to the lungs, but to a lesser extent.
Finally, the Grad-CAM of experiment 3 (fourth row of Fig \ref{fig:XR-examples}) presents the areas of interest where the segmentation procedure is carried out. In this case, the network is forced to look at the lungs, and therefore this scenario is supposed to be more realistic and more prone to generalizing as artifacts that might bias the results are somehow discarded. 

\begin{figure*}[th!]
    \setlength{\tempheight}{0.18\textheight}
    \settowidth{\tempwidth}{\includegraphics[height=\tempheight]{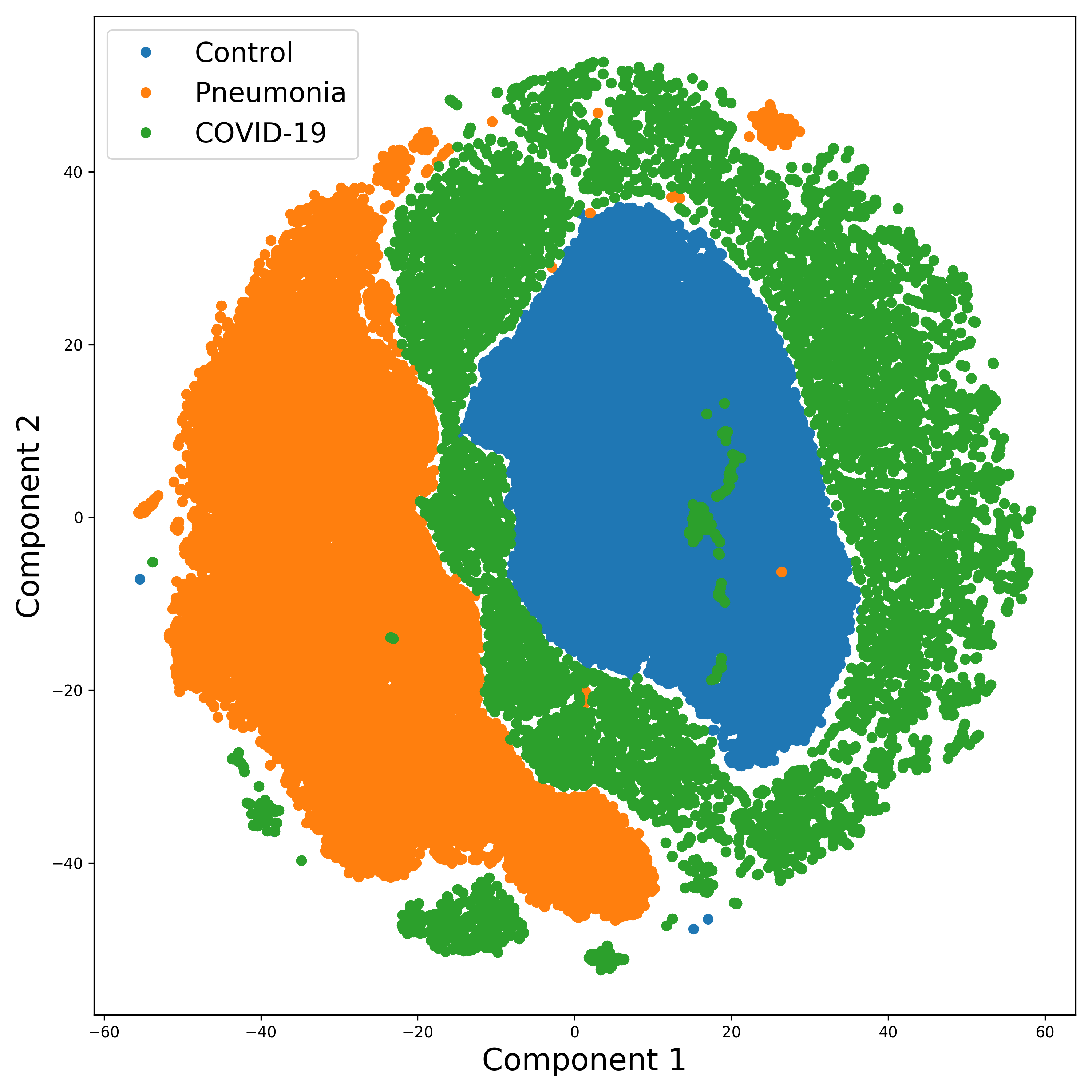}}
      \centering
      \hspace{\baselineskip}
      \columnname{Exp. 1}\hfil
      \columnname{Exp. 2}\hfil
      \columnname{Exp. 3}\\
      \rowname{Training data}
         \subfloat[]
         {
             \label{t-SNE Original Trai}
             \includegraphics[width=0.3\textwidth]{images/Exp3_OrgImages_t-SNE_training_v2.png}
         }
         \subfloat[]
         {
             \label{t-SNE Cropped Trai}
             \includegraphics[width=0.3\textwidth]{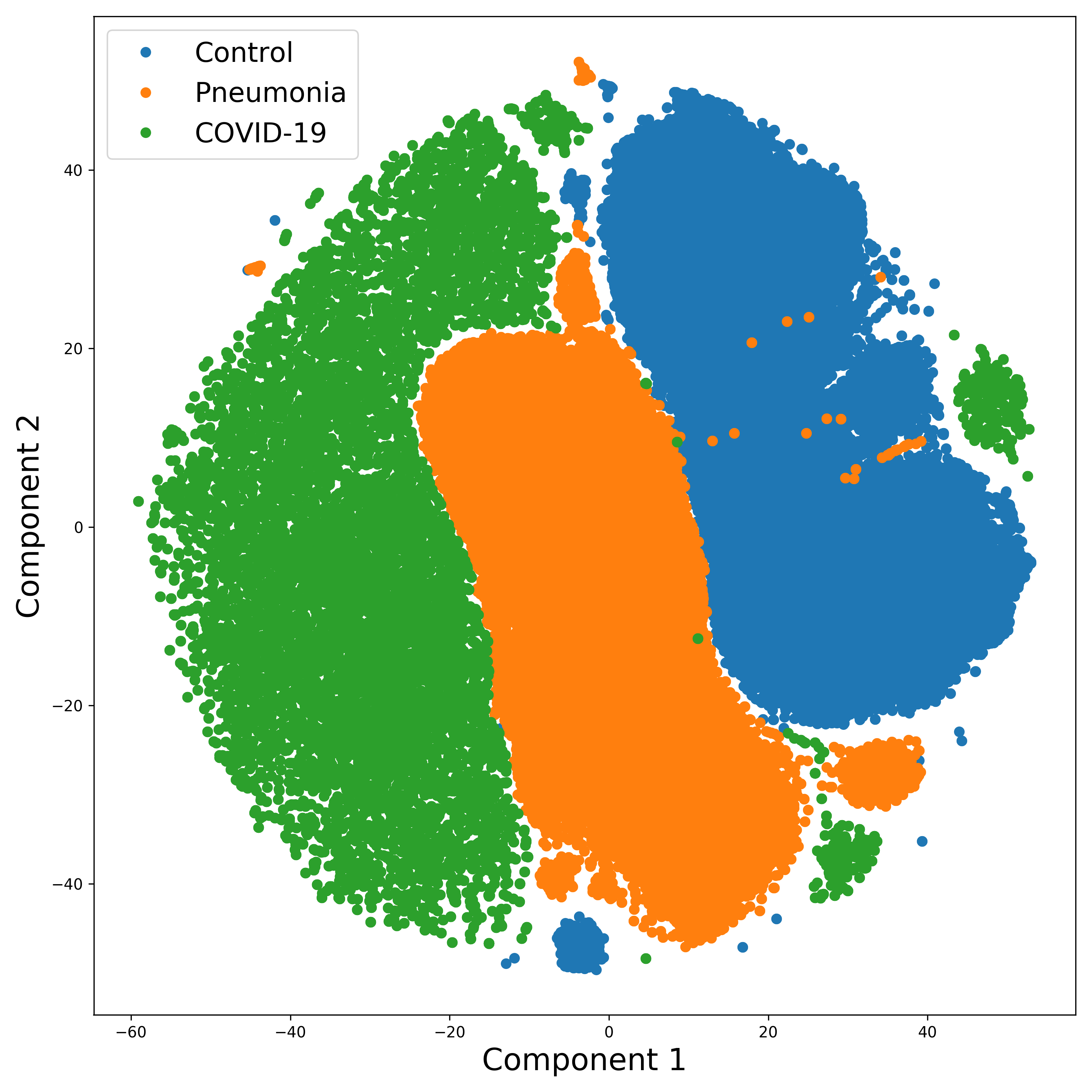}
         }
         \subfloat[]
         {
             \label{t-SNE CropSeg Trai} 
             \includegraphics[width=0.3\textwidth]{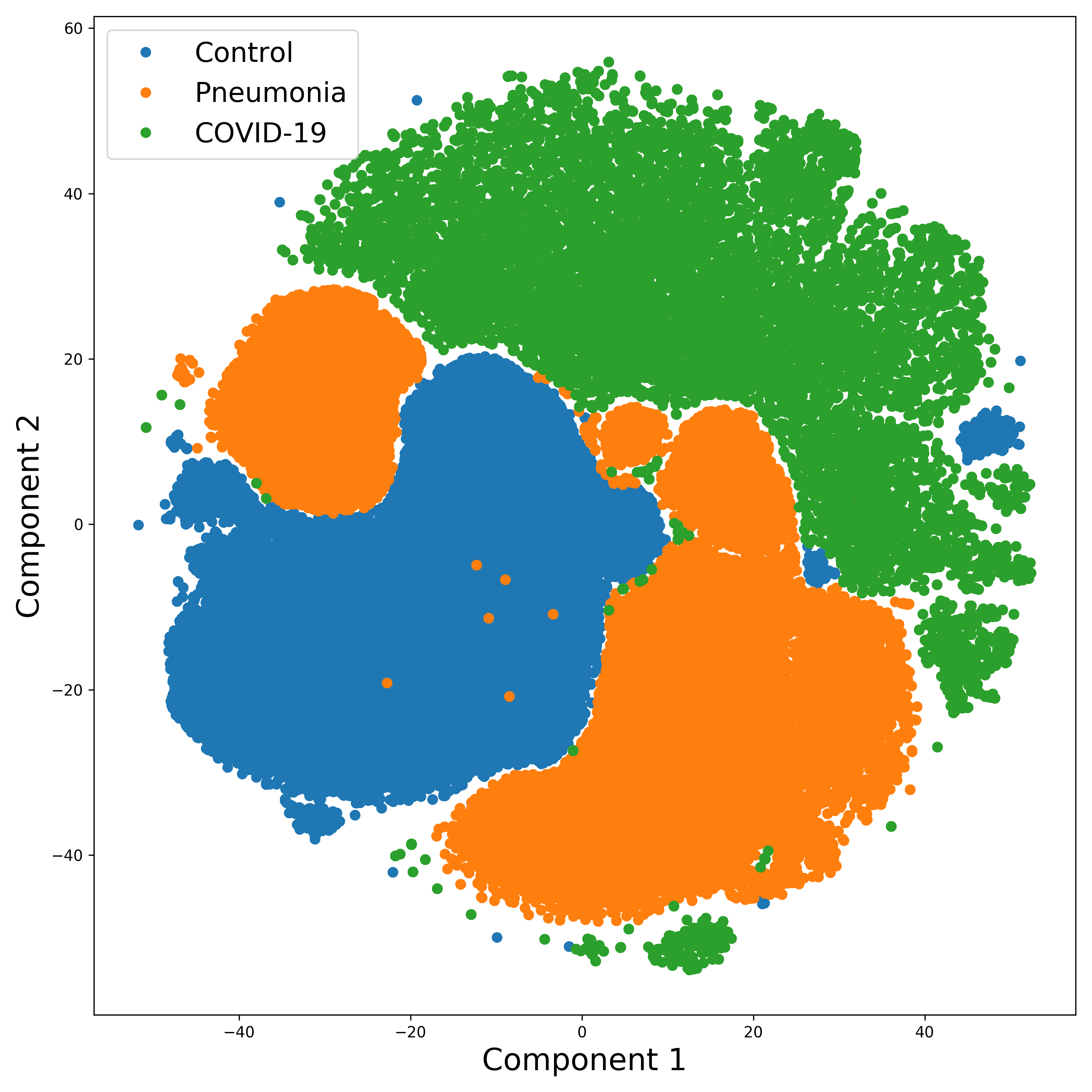}
         } 
         \\ 
        \rowname{Test data}
         \subfloat[]
         {
             \label{t-SNE Original Test}
             \includegraphics[width=0.3\textwidth]{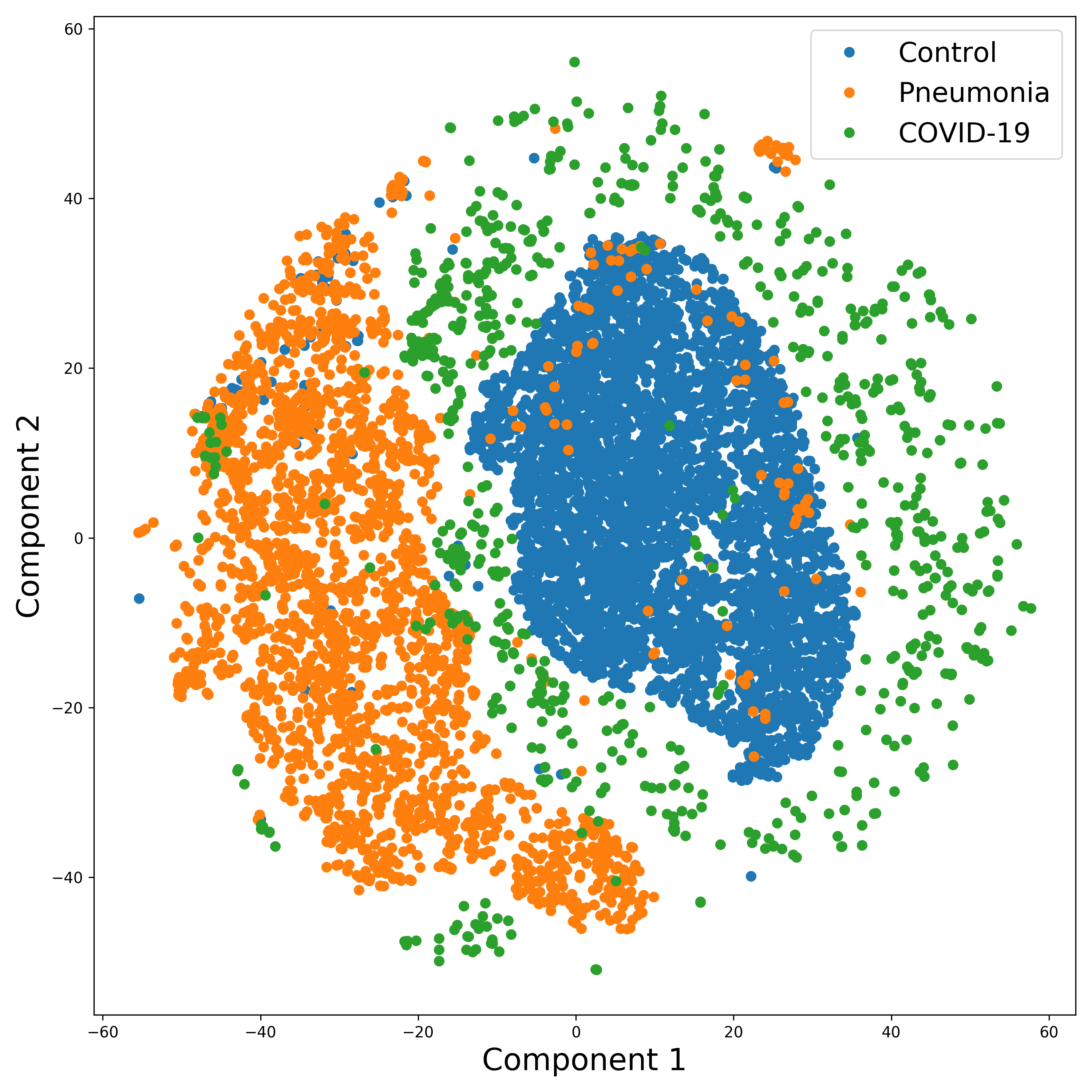}
         } 
             \subfloat[]
         {
             \label{t-SNE Cropped Test}  
             \includegraphics[width=0.3\textwidth]{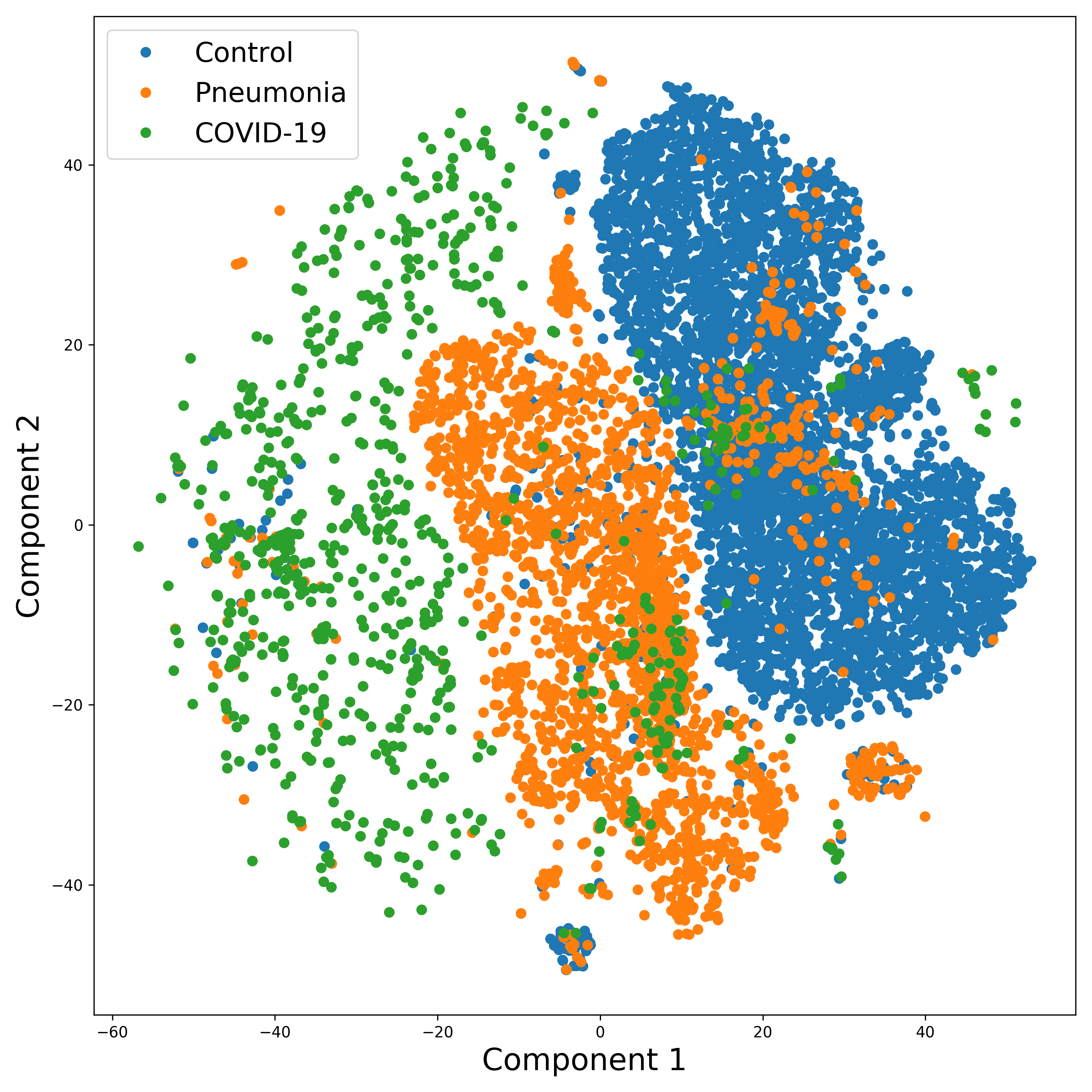}
         } 
         \subfloat[]
         {
             \label{t-SNE CropSeg Test}
             \includegraphics[width=0.3\textwidth]{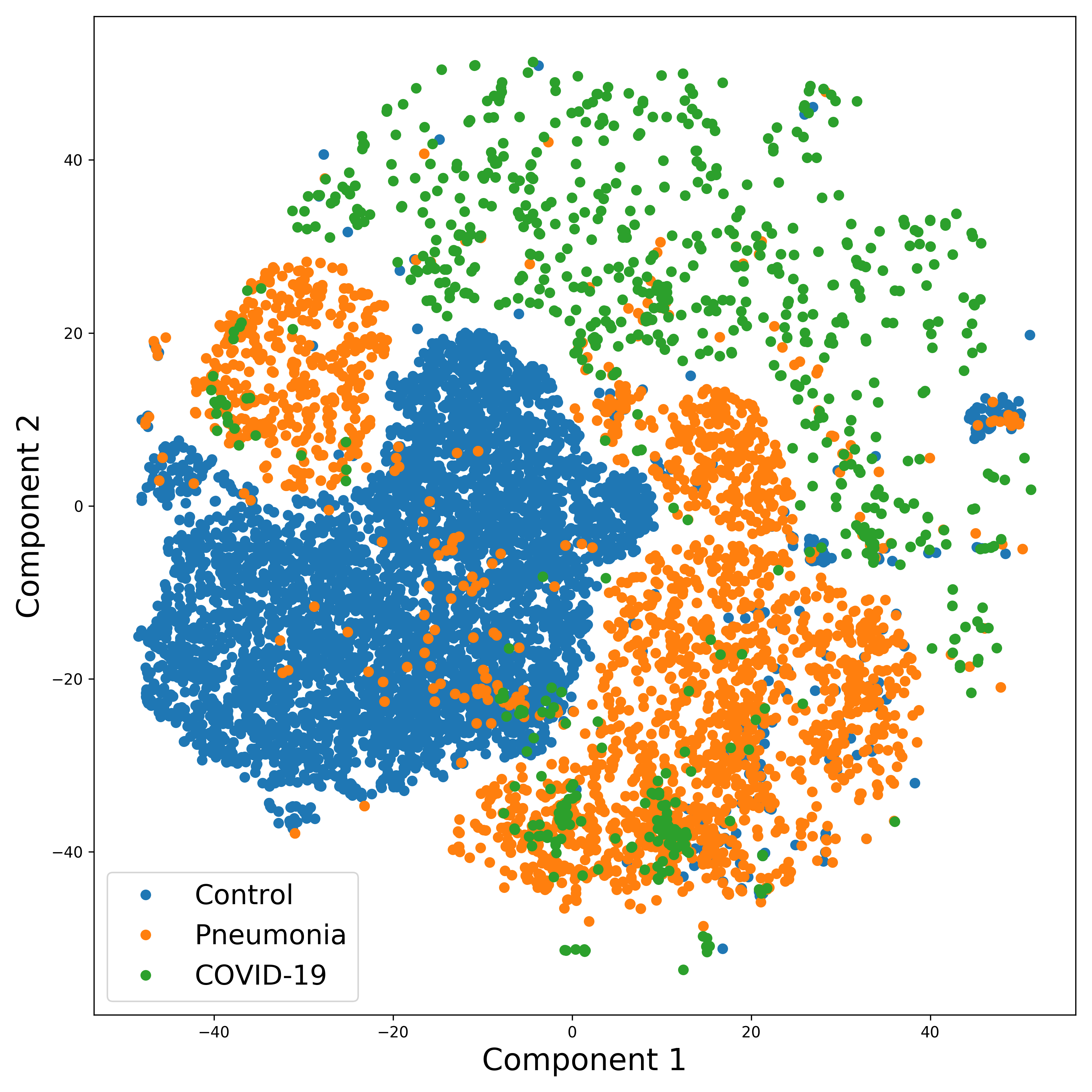}
         }
         \caption{Mapping of the high-dimensional data of the layer adjacent to the output into a two dimensional plot. \textbf{Top:} Output network embedding using t-SNE for the training data. \textbf{Bottom:} Output network embedding using t-SNE for the testing data. \textbf{Left:} Original images (experiment 1). \textbf{Center:} Cropped Images (experiment 2). \textbf{Right:} Segmented images (experiment 3). } \label{fig:t-SNE_Plots}
 \end{figure*}

On the other hand, for visualization purposes, and in order to interpret the separability capabilities of the system, a t-SNE embedding is used to project the high dimensional data of the layer adjacent to the output of the network, to a 2-dimensional space.
Results are presented in Fig. \ref{fig:t-SNE_Plots} for each of the three experiments considered in the paper.

\begin{figure*}[th!]
    \setlength{\tempheight}{0.18\textheight}
    \settowidth{\tempwidth}{\includegraphics[height=\tempheight]{images/Exp3_OrgImages_t-SNE_training_v2.png}}
      \centering
      \hspace{\baselineskip}
      \columnname{Exp. 1}\hfil
      \columnname{Exp. 2}\hfil
      \columnname{Exp. 3}\\
      \rowname{Training data}
         \subfloat[]
         {
             \label{t-SNE Original Trai 2}
             \includegraphics[width=0.3\textwidth]{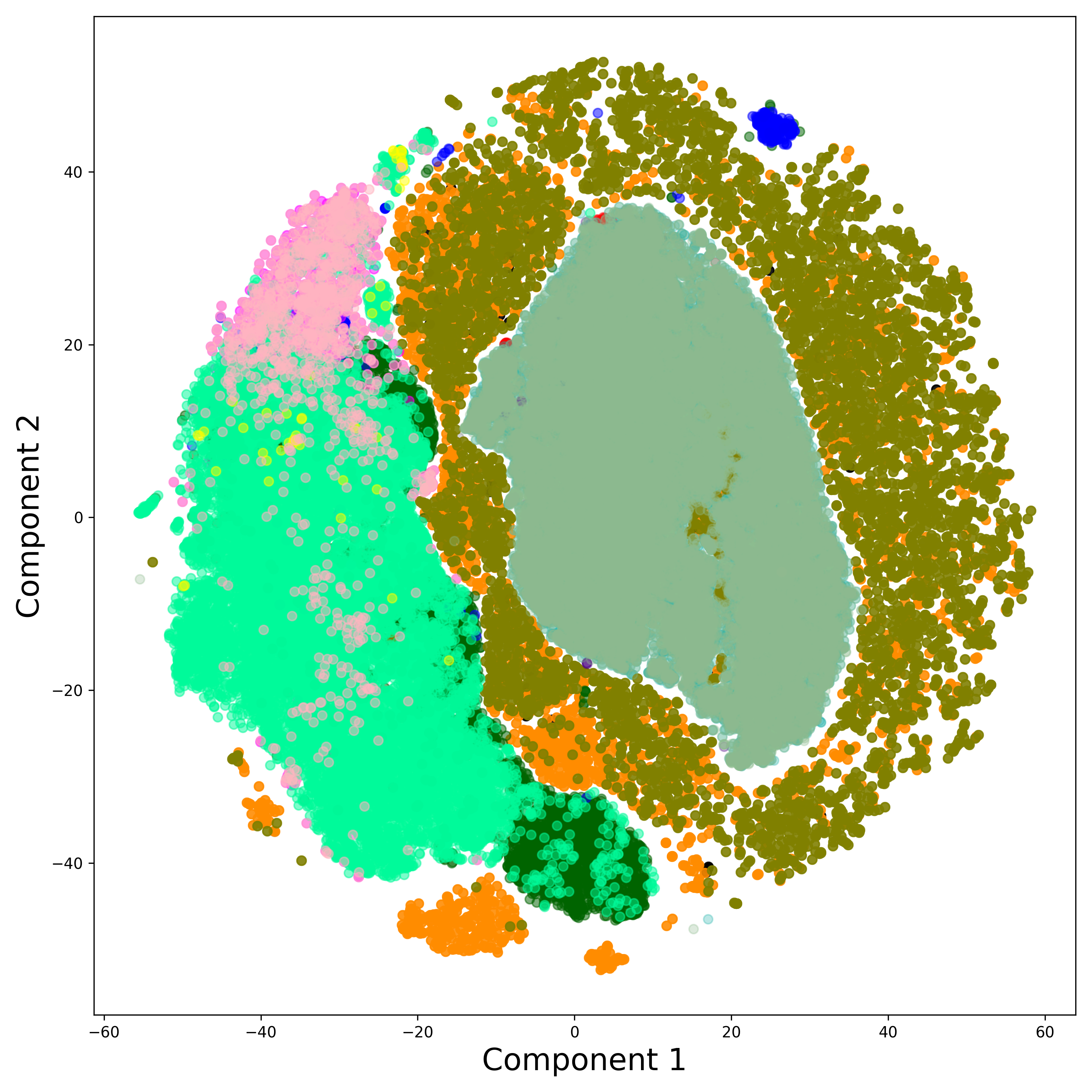}
         }
         \subfloat[]
         {
             \label{t-SNE Cropped Trai 2}
             \includegraphics[width=0.3\textwidth]{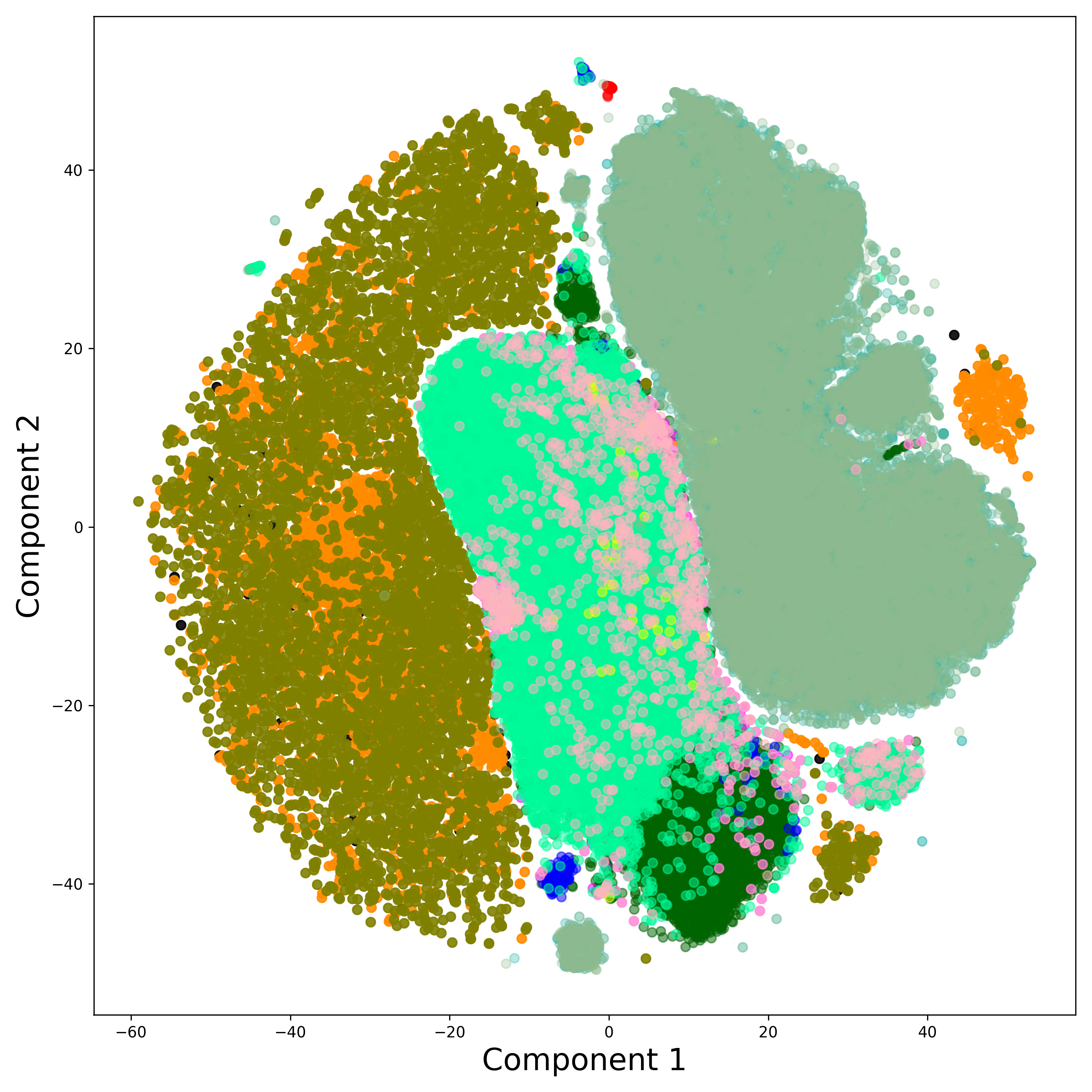}
         }
         \subfloat[]
         {
             \label{t-SNE CropSeg Trai 2} 
             \includegraphics[width=0.3\textwidth]{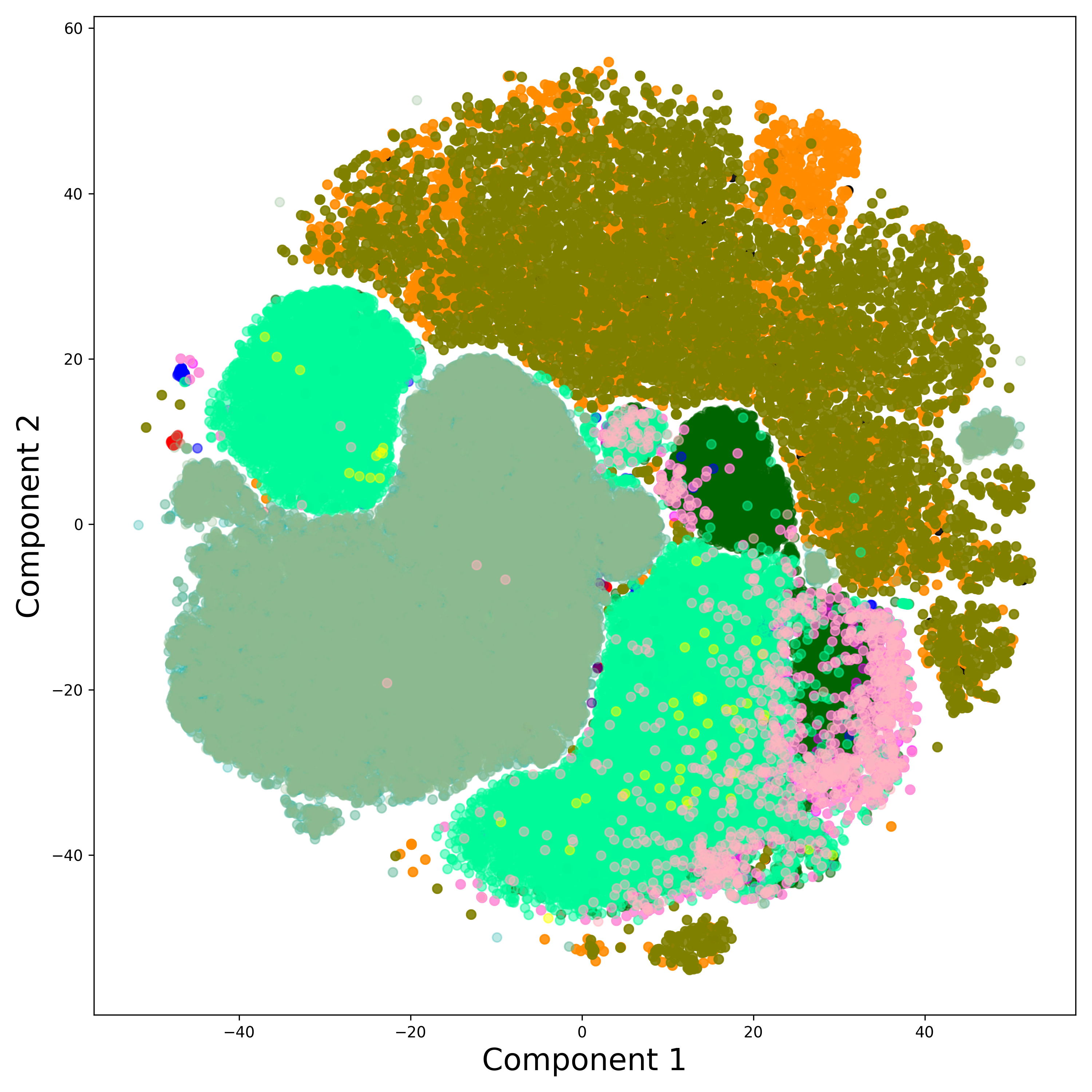}
         } 
         \\ 
        \rowname{Test data}
         \subfloat[]
         {
             \label{t-SNE Original Test 2}
             \includegraphics[width=0.3\textwidth]{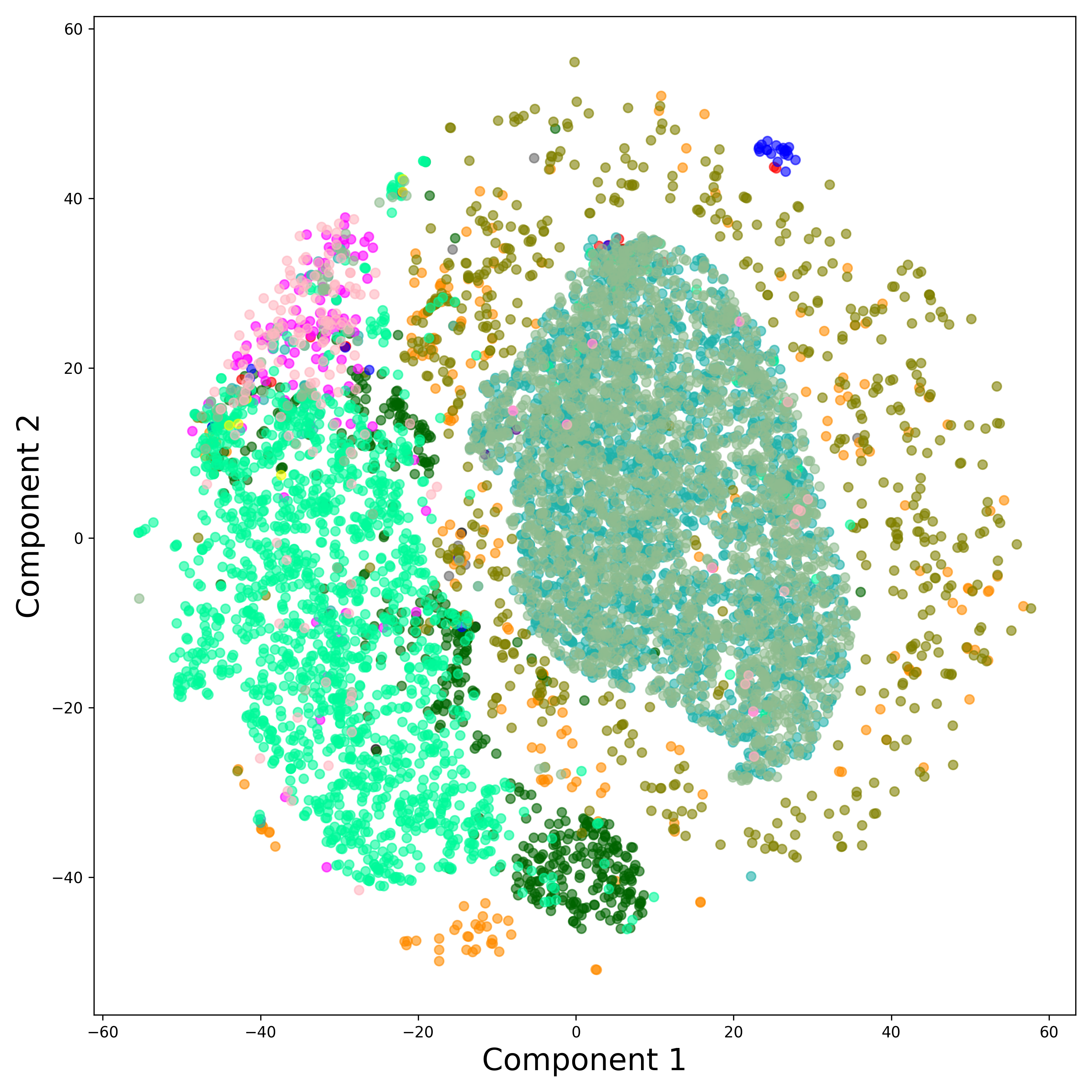}
         } 
             \subfloat[]
         {
             \label{t-SNE Cropped Test 2}  
             \includegraphics[width=0.3\textwidth]{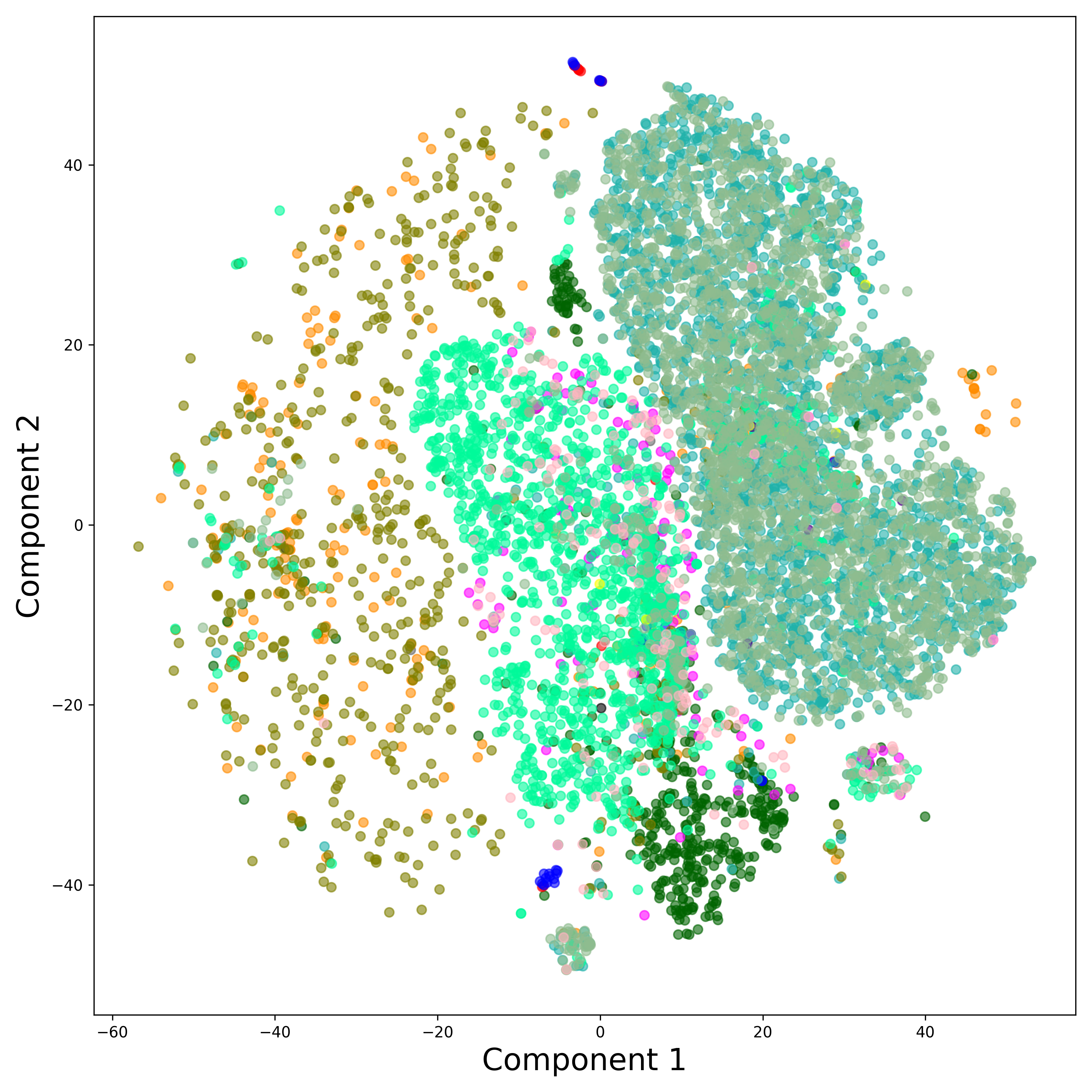}
         } 
         \subfloat[]
         {
             \label{t-SNE CropSeg Test 2}
             \includegraphics[width=0.3\textwidth]{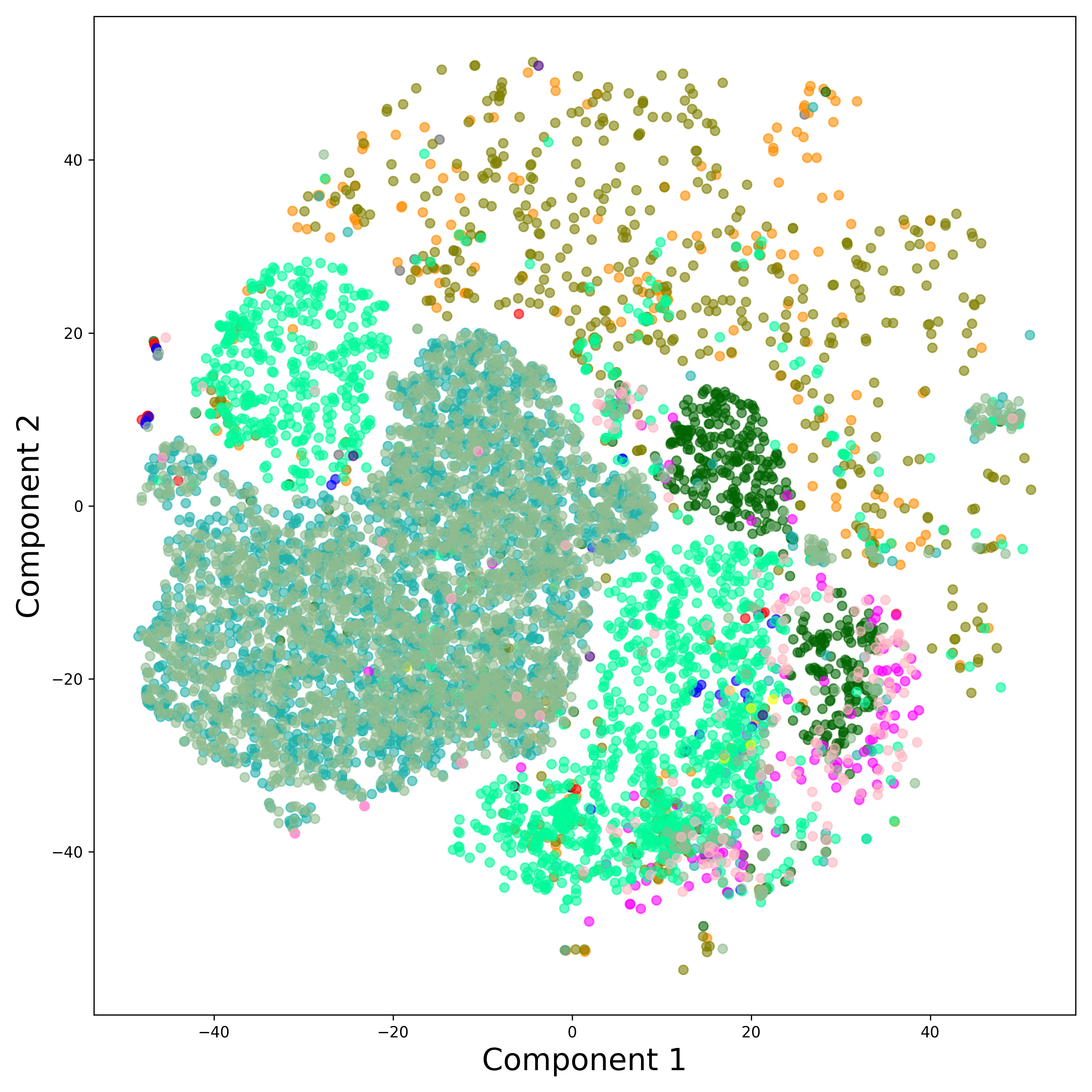}
         }
         \\
         {
             \includegraphics[scale=0.3]{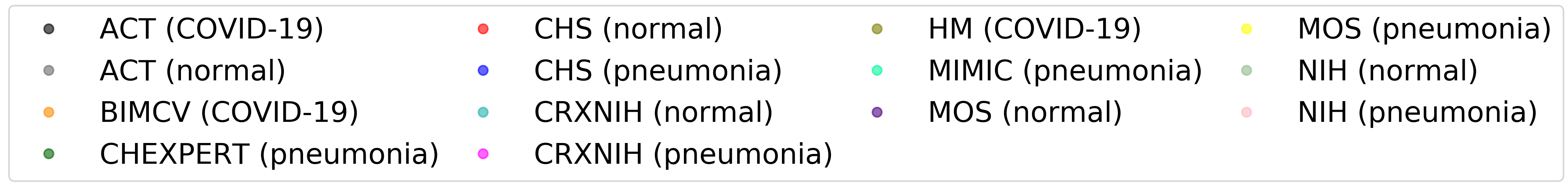}
         }
         \caption{Mapping of the high-dimensional data of the layer adjacent to the output into a two dimensional plot. \textbf{Top:} Output network embedding using t-SNE for the training data. \textbf{Bottom:} Output network embedding using t-SNE for the testing data. \textbf{Left:} Original images (experiment 1). \textbf{Center:} Cropped Images (experiment 2). \textbf{Right:} Segmented images (experiment 3). Labels correspond to data sets and classes.} \label{fig:t-SNE_Plots_v2}
 \end{figure*}

Fig. \ref{fig:t-SNE_Plots} indicates that a good separability exists for all the classes in both training and testing data, and for all experiments. The boundaries of the normal cluster are very well defined in the three experiments, whereas pneumonia and COVID-19 are more spread, overlapping with adjacent classes. 
In general terms, the t-SNE plots demonstrate the ability of the network to learn a mapping from the input data to the desired labels.  However, despite the shape differences found for the three experiments, no additional conclusions can be extracted.

\subsection{Potential variability factors affecting the system} 

There are several variability factors which might be biasing the results, namely: the projection (PA vs. AP); the technology of the detector (\textit{Computed Radiography} (CR) vs. \textit{Digital Radiography} (DX)); the gender of the patients; the age; potential specificities of the dataset; or having trained with several images per patient. 

The use of several images per patient represents a certain risk of data leak in the COVID-19 class due to its underlying imbalance. However, our initial hypothesis is that using several images per COVID-19 patient but obtained at different instants in time (with days of difference), would increase the variability of the dataset, and thus that source of bias would be disregarded.
Indeed, the evolution of the associated lesions often found in COVID-19 is considered fast, in such a manner that very different images are obtained in a time interval as short as one or two days of the evolution. 
Also, since every single exploration is framed differently, or sometimes even taken with different machines and/or projections, the potential bias is expected to be minimized.

Concerning the type of projection, and to evaluate its effectiveness, the system has been studied taking into account this potential variability factor, which is considered to be one of the most significant. 
In particular, Table \ref{tab:NumericResults2}, presents the outcomes after accounting for the influence of the XR projection (PA/AP) in the performance of the system. 
In general terms, the system demonstrates consistency with respect to the projection used, and differences are mainly attributable to smaller training and testing sets. However, significant differences are shown for projection PA in class COVID-19/experiment 3, decreasing the F1 up to $65.61$\%. The reason for the unexpected drop in performance is unknown, but likely attributable to an underrepresented class in the corpus (see Table \ref{tab:DemographicDistribution}). 

Besides, Table \ref{tab:variability_factors} shows --for the three experiments under evaluation and for the COVID-19 class-- the error distribution with respect to the sex of the patient, technology of the detector, dataset and projection. For the four variability factors enumerated, results show that the error distribution committed by the system follows --with minor deviations-- the existing proportion of the samples in the corpus. These results suggest that there is no clear bias with respect to these potential variability factors, at least for the COVID-19 class which is considered the worst case due to its underrepresentation. Similar results would be expected for control and pneumonia classes, but these results are not provided due to the lack of certain labels in some of the datasets used (see Table \ref{tab:DemographicDistribution}). 

Concerning age, the datasets used are reasonably well balanced (Table \ref{tab:DemographicDistribution}), but with a certain bias in the normal class: COVID-19 and pneumonia classes have very similar average ages, but controls have a lower mean age. 
Our assumption has been that age differences are not significantly affecting the results, but the mentioned difference might explain why the normal cluster in Fig. \ref{fig:t-SNE_Plots} is less spread than the other two. In any case, no specific age biases have been found in the errors committed by the system. 

An additional study was also carried out to evaluate the influence of potential specificities of the different datasets used to compile the corpus  (i.e. the variability of the results with respect to the datasets merged to build the corpus). This variability factor is evaluated in Fig. \ref{fig:t-SNE_Plots_v2} using different t-SNE plots (one for each experiment in a similar way than in Fig. \ref{fig:t-SNE_Plots}) but differentiating the corresponding cluster for each dataset and class. 

Results for the different datasets and classes are clearly merged or are adjacent in the same cluster. However, several datasets report a lower variability for certain classes (i.e. variability in terms of scattering). This is especially clear in Chexpert and NIH pneumonia sets, which are successfully merged with the corresponding class, but appear clearly clustered, suggesting that these datasets have certain unknown specific characteristics different to those of the complementary datasets. The model has been able to manage this aspect but is a factor to be analyzed in further studies.

\begin{table*}[!ht]
\centering
\caption{Performance measures considering the XR projection (PA/AP)}\label{tab:NumericResults2}
\begin{tabular}{@{}lclccccc@{}}
\toprule
\multirow{2}{*}{\textbf{Experiment}} & \multirow{2}{*}{\textbf{Class}} & \multicolumn{3}{c}{\textbf{PA}}  &   \multicolumn{3}{c}{\textbf{AP}}                                                                                                                            \\ \cmidrule(l){3-8} 
                                     &                                 & \textbf{PPV}     & \textbf{Recall}    & \textbf{F1}      & \textbf{PPV}                      & \textbf{Recall}                     & \textbf{F1}             \\ \midrule
\multirow{3}{*}{\textbf{Exp. 1}}     & \textit{Pneumonia}              & 91.25 $\pm$ 1.22 & 92.78 $\pm$ 1.58   & 92.00 $\pm$ 0.93 & 94.70 $\pm$ 0.79 & 96.28 $\pm$ 1.10 & 95.48 $\pm$ 0.50 \\
                                     & \textit{Control}                & 98.54 $\pm$ 0.33 & 97.83 $\pm$   0.23 & 98.18 $\pm$ 0.14 &                  97.87 $\pm$ 0.28 & 95.46 $\pm$   0.87 &   96.65 $\pm$ 0.43                              \\
                                     & \textit{COVID-19}               & 84.06 $\pm$ 3.94 & 88.91 $\pm$  2.31  & 86.33 $\pm$ 1.80 &  95.13 $\pm$ 2.46  &  97.18 $\pm$ 0.94 & 96.12 $\pm$ 1.06                                  \\ \midrule
\multirow{3}{*}{\textbf{Exp. 2}}     & \textit{Pneumonia}              & 81.77 $\pm$ 1.79 & 79.17 $\pm$ 2.38   & 80.41 $\pm$ 1.16 & 87.39 $\pm$ 1.66 & 90.78 $\pm$ 1.21 & 89.03 $\pm$ 0.71 \\
                                     & \textit{Control}   &   94.81 $\pm$ 0.46          &  95.56 $\pm$ 0.61  & 95.33 $\pm$ 0.16    & 92.79 $\pm$ 1.53  &  88.15 $\pm$  1.61                               &  90.38 $\pm$ 0.32                                                                   \\
                                     & \textit{COVID-19}               & 73.72 $\pm$ 2.37 & 68.82 $\pm$ 5.20   & 71.01 $\pm$ 2.27 & 84.96 $\pm$ 2.27                                  & 87.63 $\pm$ 2.04                                  & 86.23 $\pm$ 0.86                                  \\ \midrule
\multirow{3}{*}{\textbf{Exp. 3}}     & \textit{Pneumonia}              & 84.07 $\pm$ 1.72 & 87.19 $\pm$ 1.66   & 85.57 $\pm$ 0.53  & 87.39 $\pm$ 0.97 & 81.66 $\pm$ 1.12 & 89.47 $\pm$ 0.41 \\
                                     & \textit{Control}                & 97.88 $\pm$ 0.36 &  97.08$\pm$ 0.21   &  97.48$\pm$ 0.19 & 96.03 $\pm$ 0.81                                  & 90.65 $\pm$ 0.87                                  & 93.26 $\pm$ 0.47                                  \\
                                     & \textit{COVID-19}               & 66.68 $\pm$ 4.82 & 65.23 $\pm$ 4.73   & 65.61 $\pm$ 1.05 & 81.82 $\pm$ 3.07                                  &  83.62 $\pm$ 2.14                                 & 82.65 $\pm$ 1.28                                   \\ \bottomrule
\end{tabular}

\end{table*}

\begin{table}[ht!]
    \centering
    \caption{Percentage of testing samples and error distribution with respect to several potential variability factors for the COVID-19 class. (\% in hits represents the percentage of samples of every factor under analysis in the correctly predicted set.) }\label{tab:variability_factors}
    \begin{tabular}{lccccc}
    \toprule
    \multirow{2}{*}{\textbf{Factor}} & \multirow{2}{*}{\textbf{Types}} & \multirow{2}{*}{\textbf{\% in test}} & \multicolumn{3}{c}{\textbf{\% in hits}}  \\ \cmidrule(l){4-6}
         &   &  & \textbf{Exp. 1} & \textbf{Exp. 2} & \textbf{Exp. 3} \\ \hline
    \multirow{2}{*}{\textbf{Projection}} & AP   &  79  & 80.0  & 82.6   & 82.7  \\
                                         & PA   &  21  & 20.0  & 17.4   & 17.3  \\ \hline
    \multirow{2}{*}{\textbf{Sensor}}     & DX   &  22  & 22.0  & 23.3   & 23.6  \\
                                         & CR   &  78  & 78.0  & 76.7   & 76.4  \\ \hline
    \multirow{2}{*}{\textbf{Sex}}        & M    &  64  & 64.0  & 65.4   & 65.2   \\
                                         & F    &  36  & 36.0  & 34.6   & 34.8   \\ \hline
    \multirow{3}{*}{\textbf{DB}}         & BMICV    &  30  & 28.7  & 26.6   & 26.6  \\  
                                         & HM    & 69   & 71.0  & 72.7   & 73.1  \\
                                         & ACT    &  1  & 0.3  & 0.7   & 0.3  \\
      \bottomrule
    \end{tabular}
    
\end{table}

\section{Discussion and Conclusions}
\label{sec:disscon} 

This study evaluates a deep learning model for the detection of COVID-19 from RX images. The paper provides additional evidence to the state of the art, supporting the potentiality of deep learning techniques to accurately categorize XR images corresponding to control, pneumonia, and COVID-19 patients (Fig. \ref{fig:XR-examples}). These three classes were chosen under the assumption that they can support clinicians on making better decisions, establishing potential differential strategies to handle patients depending on their cause of infection \cite{wang2020covid}.
However, the main goal of the paper was not to demonstrate the suitability of the deep learning for categorizing XR images, but to make a thoughtful evaluation of the results and of different preprocessing approaches, searching for better explainability and/or interpretability of the results, while providing evidence of potential effects that might bias results.   

The model relies on the COVID-Net network, which has served as a basis for the development of a more refined architecture. This network has been chosen due to its tailored characteristics and given the previous good results reported by other researchers. The COVID-Net was trained with a corpus compiled using data gathered from different sources: the control and pneumonia classes --with $49,983$ and $24,114$ samples respectively-- were collected from the ACT, Chinaset, Montgomery, CRX8, CheXpert and MIMIC datasets; and the COVID-19 class was collected from the information available at the BIMCV, ACT, and HM Hospitales datasets.  

Although the COVID-19 class only contains $8,573$ chest RX images, the developers of the data sources are continuously adding new cases to the respective repositories, so the number of samples is expected to grow in the future. 
Despite the unbalance of the COVID-19 class, up to date, and to the authors' knowledge, this is the largest compilation of COVID-19, images based on open repositories.
Despite that, the number of COVID-19 RX images is still considered small in comparison to the other two classes, and therefore, it was necessary to compensate for the class imbalance by modifying the network architecture, including regularization components in the last two dense layers.
To this end, a weighted categorical cross-entropy loss function was used to compensate for this effect.
Likewise, data augmentation techniques were used for pneumonia and COVID-19 classes to automatically generate more samples for these two underrepresented classes.

We stand on the fact that automatic diagnosis is much more than a classification exercise, meaning that many factors have to be had in mind to bring these techniques to the clinical practice. 
To this respect, there is a classic assumption in the literature that the associated heat maps --calculated with techniques such as Grad-CAM-- provide a clinical interpretation of the results, which is unclear in practice. In light of the results shown in the heat maps depicted in Fig. \ref{fig:XR-examples}, we show that experiment 1 must be carefully interpreted. 
Despite the high-performance metrics obtained in experiment 1, the significant areas identified by the network are pointing towards certain areas with no clear interest for the diagnosis, such as  corners of the images, the sternum, clavicles, etc. From a clinical point of view, this is clearly biasing the results. 
It means that other approaches are necessary to force the network to focus on the lungs area. To this respect, we have developed and compared the results with two preprocessing approaches based on cropping the images and segmenting the lungs area (experiment 2 and experiment 3). 
Again, given the heat maps corresponding to experiment 2, we also see similar explainability problems to those enumerated for experiment 1. 
Reducing the area of interest to that proposed in experiment 2 significantly decreases the performance of the system due to the removal of the metadata that usually appear in the top left and/or right corner, and to the removal of areas which are of interest to categorize the images but have no interest from the diagnosis point of view. 
However, while comparing experiment 2 and 3, performance results improve in the third approach, which focuses on the same region of interest but with a mask that forces the network to see only the lungs. Thus, results obtained in experiments 2 and 3 suggest that eliminating the needless features extracted from the background or non-related regions improves the results. 
Besides, the third approach (experiment 3), provides more explainable and interpretative results, with the network focusing its attention only in the area of interest for the disease. The gain in explainability of the last method is still at the cost of a lower accuracy with respect to experiment 1, but the improvement in explainability and interpretability are considered critical to translate these techniques to the clinical setting. Despite the decrease in performance, the proposed method in experiment 3 has provided promising results, with an Acc of $91.53\%$, BAcc of $87.6$, GMR of $87.37\%$ and AUC of $0.97$. 

Performance results obtained are in line with those presented in \cite{wang2020covid}, which reports sensitivities of $95$, $94$ and $91$ for control, pneumonia and COVID-19 classes respectively --also modelling with the COVID-Net in a scenario similar to the one in experiment 1--, but training with a much smaller corpus of $358$ RX images from $266$ COVID-19 patients, $8,066$ controls, and $5,538$ RX images belonging to patients with different types of pneumonia. 

The paper also critically evaluates the effect of several variability factors that might compromise the performance of the network. 
For instance, the effect of the projection (PA/AP) was evaluated by retraining the network and checking the outcomes.
This effect is important, given that PA projections are often practised in erect positions to better observe the pulmonary ways, and as such, are expected to be examined in healthy or slightly affected patients.  
In contrast, AP projections are often preferred for patients confined in bed, and as such are expected to be practised in the most severe cases. Since AP projections are common in COVID-19 patients, blood is expected to flow more to lungs’ apices than when standing; thus, not considering this variability factor into account may result in a misdiagnosis of pulmonary congestion \cite{burlacu2020curbing}. 
Indeed, the obtained results have highlighted the importance of taking into account this factor when designing the training corpus, as PPV have decreased for PA projections in our experiments with the COVID-19 images. This is probably due to an underrepresentation of this class (Table \ref{tab:NumericResults2}), which would require a further specific analysis when designing future corpora. 

On the other hand, results have shown that the error distribution for the COVID-19 class follows a similar proportion than the percentage of images available in the corpus while categorizing by gender, the technology of the detector, the projection and/or the dataset. 
These results suggest no significant bias with respect to these potential variability factors, at least for the COVID-19 class, which is the less represented one. 

An analysis of how the clusters of classes were distributed is also presented in Fig. \ref{fig:t-SNE_Plots}, demonstrating how well each class is differentiated. These plots help to identify existing overlap among classes (especially that present between pneumonia and COVID-19, and to a lesser extent between controls and pneumonia).
Similarly, since the corpus used to train the network was built around several datasets, a new set of t-SNE plots was produced, but differentiating according to each of the subsets that were used for training (Fig. \ref{fig:t-SNE_Plots_v2}).
This test served to evaluate the influence of potential specific characteristics of each dataset in the training procedure and, hence, possible sources of confusion that arise due to particularities of the corpora that are tested.
The plots suggest that in general terms the different datasets are correctly merged together, but with some exceptions. This fact suggests that there might be certain unknown characteristics in the datasets used, which cluster the images belonging to the same dataset together. 

The COVID-Net has also demonstrated being a good starting point for the characterization of the disease. Indeed, the outcomes of the paper suggest the possibility to automatically identifying the lung lesions associated with a COVID-19 infection (see Fig.\ref{fig:XR-examples}) by analyzing the Grad-CAM mappings of experiment 3, providing an explainable justification about the way the network works. However, the interpretation of the heat maps obtained for the control class must be carried out carefully. Whereas the areas of significant interest for pneumonia and COVID-19 classes are supposed to point to potential lesions (i.e. with higher density and/or with different textures in contrast to controls), the areas of significant interest for the classification in the control group are supposed to correspond to a sort of complement, potentially highlighting less dense areas. Thus, not meaning the presence of any kind of lesion in the lungs. 
 
Likewise, and in comparison to the performance achieved by a human evaluator differentiating pneumonia from COVID-19, the system developed in the third experiment attains comparable results.
Indeed, in \cite{bai2020performance} the ability of seven radiologists to correctly differentiate pneumonia and COVID-19 from XR images was put into test. 
The results indicated that the radiologists achieved sensitivities ranging from $97\%$ to $70\%$ (mean $80\%$), and specificities ranging from $7\%$ to $100\%$ (mean $70\%$). These results suggest a potential use in a supervised clinical environment. 

COVID-19 is still a new disease and much remains to be studied. 
The use of deep learning techniques would potentially help to understand the mechanisms on how the SARS-CoV2 attacks the lungs and alveoli, and how it evolves during the different stages of the disease.
Despite there is some empirical evidence on the evolution of COVID-19 --based on observations made by radiologists \cite{pan2020imaging}--, the employment of automatic techniques based on machine learning would help to analyze data massively, to guide research onto certain paths or to extract conclusions faster. But more interpretable and explainable methods are required to go one step forward. 

Inline to the previous comment, and based on the empirical evidence respecting the evolution of the disease, it has been stated that during early stages of the disease, ground-glass shadows, pulmonary consolidation and nodules, and local consolidation in the centre with peripheral ground-glass density are often observed, but once the disease evolves the consolidations reduce their density resembling a ground-glass opacity, that can derive in a "white lung" if the disease worsens or in a minimization of the opacities if the course of the disease improves \cite{pan2020imaging}. 
In this manner, if any of these characteristic behaviours are automatically identified, it would be possible to stratify the stage of the disorder according to its severity.
Not only that but computing the extent of the ground-glass opacities or densities would also be useful to assess the severity of the infection or to evaluate the evolution of the disease. 
The assessment of the infection extent has been previously tested in other CT studies of COVID-19 \cite{yang2020chest}, but using manual procedures based on observation of the images.

Solutions like the one discussed in this paper, are intended to support a much faster diagnosis and to alleviate the workload of radiologists and specialists, but not to substitute their assessment. A rigorous validation would open the door to integrating these algorithms in desktop applications or cloud servers for its use in the clinic environment. Thus, its use, maintenance and update would be simple and cost-effective, and would reduce healthcare costs, improve the accuracy of the diagnosis and the response time \cite{topol2019deep}. In any case, the deployment of these algorithms is not exempt from controversies: hosting the AI models in a cloud service would entail the upload of the images which might be subject to national and/or international regulations and constraints to ensure privacy \cite{he2019practical}.

\bibliographystyle{IEEEtran}
\balance
\bibliography{References}

\end{document}